\documentclass[aps,superscriptaddress,longbibliography,showkeys]{revtex4-2}
\usepackage{orcidlink} 
\usepackage{amsmath,amssymb,bbm}
\usepackage{physics}
\usepackage{hyperref}
\usepackage{graphicx}
\usepackage{xcolor}
\usepackage{listings}
\usepackage{dsfont}
\usepackage{amsmath}
\usepackage{amssymb}
\usepackage{graphicx}
\usepackage{babel}
\usepackage{bbm}

\hypersetup{colorlinks=true, linkcolor=blue!60!black, urlcolor=blue, citecolor=blue!60!black}

\newcommand{\pkg}[1]{\texttt{#1}}

\begin{document}


\title{Tensor network solvers for ultra-large tight-binding Hamiltonians: \\ algorithms and applications}

\author{Tiago V. C. Ant\~ao\,\orcidlink{0000-0003-3622-2513}}
\thanks{These authors contributed equally.}
\affiliation{Department of Applied Physics, Aalto University, Espoo, Finland}

\author{Anouar Moustaj\,\orcidlink{0000-0002-9844-2987}}
\thanks{These authors contributed equally.}
\affiliation{Department of Applied Physics, Aalto University, Espoo, Finland}

\author{Yitao Sun\,\orcidlink{0009-0002-9479-7147}}
\thanks{These authors contributed equally.}
\affiliation{Department of Applied Physics, Aalto University, Espoo, Finland}

\author{Jose L. Lado\,\orcidlink{0000-0002-9916-1589}}
\affiliation{Department of Applied Physics, Aalto University, Espoo, Finland}

\date{\today}

\begin{abstract}
Understanding quantum materials at meso and even macroscopic scales
requires tight-binding calculations on system sizes where explicit matrix representations become prohibitively costly.
This represents a major bottleneck to
rationalize phenomena in moir\'e and super-moir\'e heterostructures and quasicrystals.
Here, we present a unified tensor-network methodology to solve
tight-binding problems at exceptionally large scales,
by mapping a system of $N = 2^L$ sites onto 
a many-body problem of $L$ pseudospin sites,
which is subsequently solved with tensor network algorithms. For Hamiltonians with compressible real-space structure, 
the tensor network
bond dimension remains modest, typically of order a few tens, independent of $N$.
Tensor network representations of arbitrary hopping functions including long-range, spatially modulated, and twisted-layer couplings are built with quantics tensor cross interpolation, and all physical observables are evaluated entirely with tensor network algebra without explicit matrix storage or diagonalization.
We demonstrate applications to spectral functions, momentum-space spectra via the tensor-network quantum Fourier transform, real-space topological invariants, real-time dynamics, correlation-induced symmetry breaking with self-consistent mean-field calculations,
non-Hermitian phenomena, 
and excitonic many-body physics. Our
methodology enables routinely solving
systems with billions of sites, by leveraging
the tensor network compressibility of
real-space structures, and establishing a flexible framework
to study quantum matter at ultra-large length scales.
The methodology is implemented in the open-source Julia package \pkg{TensorBinding.jl}.

\end{abstract}

\maketitle

\section{Introduction}

A growing class of problems in condensed-matter physics requires computationally understanding quantum phenomena that emerge only on ultra-large real-space scales. In these settings, the relevant physics is not solely controlled by the microscopic unit cell, but rather by long-wavelength modulations, 
aperiodic order, and emergent superstructures \cite{kennes2021,balents2020,lai2025}. Moir\'e superlattices formed by stacked van der Waals layers exhibit unit cells of thousands to millions of atoms. Quasiperiodic systems, including quasicrystals, Aubry-Andr\'e-Harper models, and fractals, lack translational symmetry entirely. As a result, computing their modulated topological and spectral properties requires real-space models large enough to suppress finite-size artifacts \cite{antao2026,aubry1980,hofstadter1976,gonalves2022,gonalves2023,duncan2020,lado2019,khosravian2021,uri2023,tsang2024} with correlated electronic structure that only becomes apparent on length scales far exceeding those underlying ordered formations of the constituent crystals \cite{sun2025,cao2018insulator,cao2018sc,bistritzer2011,andrei2020,andrei2021,nuckolls2024,grover2022,li2024,hao2024,xia2025,yu2019}. 
This limitation not only applies to static properties, but also becomes especially severe when computing out-of-equilibrium properties and
excitations \cite{rudner2020,delatorre2021,onida2002,yuan2022}. Driven systems under time-dependent fields require real-time propagation of large lattice models, a regime where the cost of diagonalization-based approaches compounds at every time step \cite{oka2019}, and exciton binding in moir\'e bilayers involves electron-hole correlations that can be modulated across the entire structure of a material sample, demanding two-particle Hilbert spaces that are much larger than the single-particle one \cite{moustaj2026excitons,tran2019,wu2018,herrera2025}. In the same way, non-Hermitian systems with spatially modulated loss and gain \cite{gong2018,song2019}, or non-reciprocal hoppings require spectral and dynamical calculations whose real-space structure cannot be captured at small system sizes \cite{ashida2020,bergholtz2021,sun2026nonhermitian,yao2018,okuma2020,helbig2020}. In each of these settings, the required site counts fall far beyond what conventional diagonalization or sparse-matrix methods can sustain for the full suite of spectral, dynamical, and topological observables.

This is due to the fact that conventional tight-binding computational methodologies \cite{groth2014,pythtb,tbmodels,pyqula,joao2020,yuan2022,quantica}, despite proving to be efficient for small to moderate system sizes, rely on an explicit matrix representation of the Hamiltonian, either dense or sparse, which becomes a bottleneck for very large real-space systems on the order of billions of sites due to the polynomial scaling of memory with system size.
This limitation imposes severe constraints on the ability to compute multi-scale resolutions of physical observables or even evaluate such quantities over all sites in a dense form. Tensor networks \cite{fishman2022,hauschild2018,schollwock2005,schollwock2011,paeckel2019} and density matrix renormalization group \cite{white1992} methods, on the other hand, allow one to deal with exceptionally large many-body Hilbert spaces by exploiting low-rank tensor structure, enabling computations that are intractable with explicit matrices. While tensor-network methods have been extensively used in many-body physics, their use for single-particle phenomena remained unexplored until recently\cite{jcjmp2025}. Specifically, this new direction has been enabled by the development of quantics tensor cross interpolation, a quantum-inspired active learning algorithm that allows for a re-purposing of the tensor-network methodology to encode exponentially classical functions. This strategy has been demonstrated to solve partial differential equations \cite{niedermeier2026,boucomas2025}, fluid dynamics \cite{peddinti2024,gourianov2025}, wavefunctions in computational chemistry \cite{li2026,nakatani2013,szalay2015,chan2016}, many-body diagrammatics and impurity solvers \cite{nunezfernandez2022,erpenbeck2023,murray2024}, machine learning \cite{stoudenmire2016,han2018,dilip2022}, and quantum computation \cite{zhou2020,huang2021,farrelly2021,farrelly2022}, suggesting that the same compression ideas can be brought to bear on tight-binding problems. However, no unified framework exists yet that enables one to flexibly solve tight-binding problems with the scalability of tensor-network methods.

\begin{figure}[!t]
    \centering
    \includegraphics[width=0.65\linewidth]{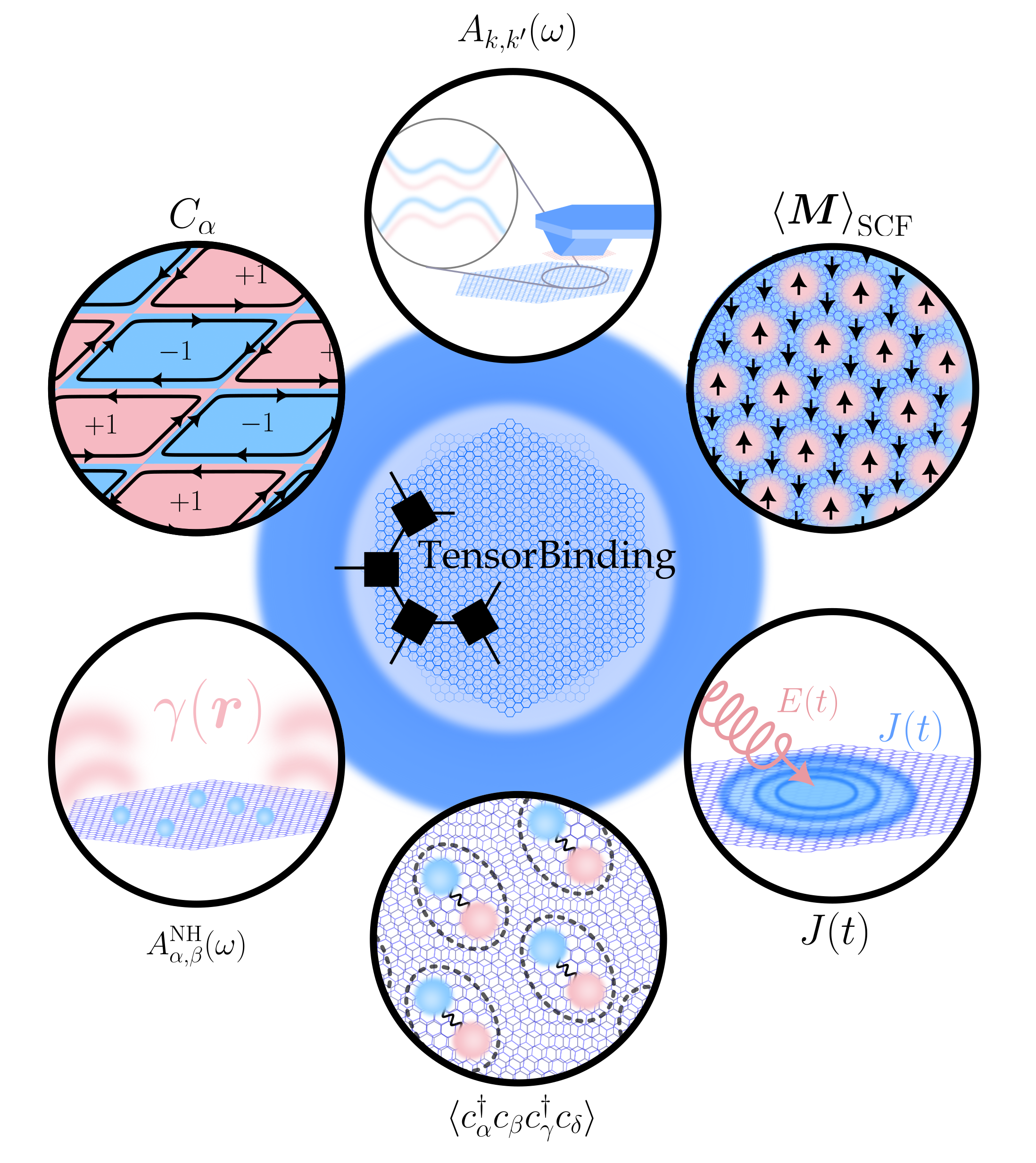}
    \caption{Overview of the \pkg{TensorBinding.jl} framework and its current range of applications.
    The panels illustrate six classes of physical observable accessible within the framework:
    the momentum-resolved spectral function $A_{k,k'}(\omega)$ (top),
    the real-space Chern marker $C_\alpha$ revealing coexisting topological phases (left),
    self-consistent mean-field correlated phases, such as the magnetization $\langle\mathbf{M}\rangle_{\rm{SCF}}$ (right),
    time-dependent observables, such as the bond current $J(t)$ under a driven electric field $E(t)$ (bottom right),
    two-particle correlated states $\langle c_\alpha^\dagger c_\beta c_\gamma^\dagger c_\delta\rangle$, such as excitons (bottom center),
    and non-hermitian spectral function $A^{NH}_{\alpha,\beta}$ for systems with modulated losses $\gamma(\mathbf{r})$ (bottom left).}
    \label{fig:schematic}
\end{figure}

Here, we aim to fill this gap by presenting a unified tensor-network framework to solve a variety of tight-binding based problems, some of which are illustrated in Fig.~\ref{fig:schematic}, for exceptionally large system sizes. The key strategy relies on mapping a tight-binding Hamiltonian on $N = 2^L$ sites to the many-body space of a quantum spin $S=1/2$ chain of $L$ pseudospins, as illustrated in Fig.~\ref{fig:mapping_benchmarks}(a,b). When compressed as a tensor network, this encoding \cite{khoromskij2011,oseledets2011} routinely keeps the bond dimension on the order of a few tens for a broad class of physically relevant models, independently of $N$. Arbitrary two-body hopping functions, including long-range, spatially modulated, and inter-layer couplings are then compressed into tensor networks via quantics tensor cross interpolation \cite{ritter2024,nunez2025,jeannin2024}, avoiding explicit construction of the full sparse matrix. 
All subsequent operations
and observables
can be evaluated entirely within 
tensor network algebra, bypassing explicit diagonalization and making systems of billions accessible for compressible models. These operations may include $k$-space computation of spectral functions via the Quantum Fourier Transform, real-space topological marker operators, self-consistent mean-field calculations, or even excitonic properties, among other observables, as overviewed in Fig.~\ref{fig:schematic}. This approach is implemented in \pkg{TensorBinding.jl}, an open-source Julia package built on the ITensors ecosystem \cite{fishman2022}, 
which unifies and significantly improves upon the algorithms of recent demonstrations \cite{sun2025,moustaj2025momentum,fumega2025,antao2026,moustaj2026excitons,sun2026nonhermitian},
providing a flexible computational platform for future development of tight-binding tensor network methods.

The following sections describe the tensor-network methods underpinning a wide set of observables
that can be computed with the tensor network tight binding
methodology. In Section \ref{sec:tn_rep} we survey the main algorithmic components of the tensor-network methods used across the package, with an emphasis on the construction of generic tight-binding Hamiltonians. In Section \ref{sec:single_particle}  we demonstrate an algorithm to compute topology and band structures for modulated topological models, showing that band structures can be obtained for a wide variety of complex Hamiltonians even with non-trivial topological properties. We additionally present alternative
algorithms applicable to non-Hermitian models, in particular with a modulated lossy tight-binding Hamiltonian and compute its dynamical properties, showing how losses affect the local occupation number over time. In Section \ref{sec:correlated}  we show the applicability of this method to interacting problems,
in particular demonstrating self-consistent symmetry-broken magnetic orders,
and excitonic spectral functions. In Section \ref{sec:outlook}  we provide an outlook on potential future
developments of our methodology, some of them under active development in
\pkg{TensorBinding.jl}. Finally Section \ref{sec:conclusion}  offers concluding remarks on the application range and limitations of the framework.

\section{Tensor-Network representation of tight-binding problems}
\subsection{Representation of Hamiltonians as Matrix Product Operators}
\label{sec:tn_rep}
\begin{figure}[!t]
  \centering
  \includegraphics[]{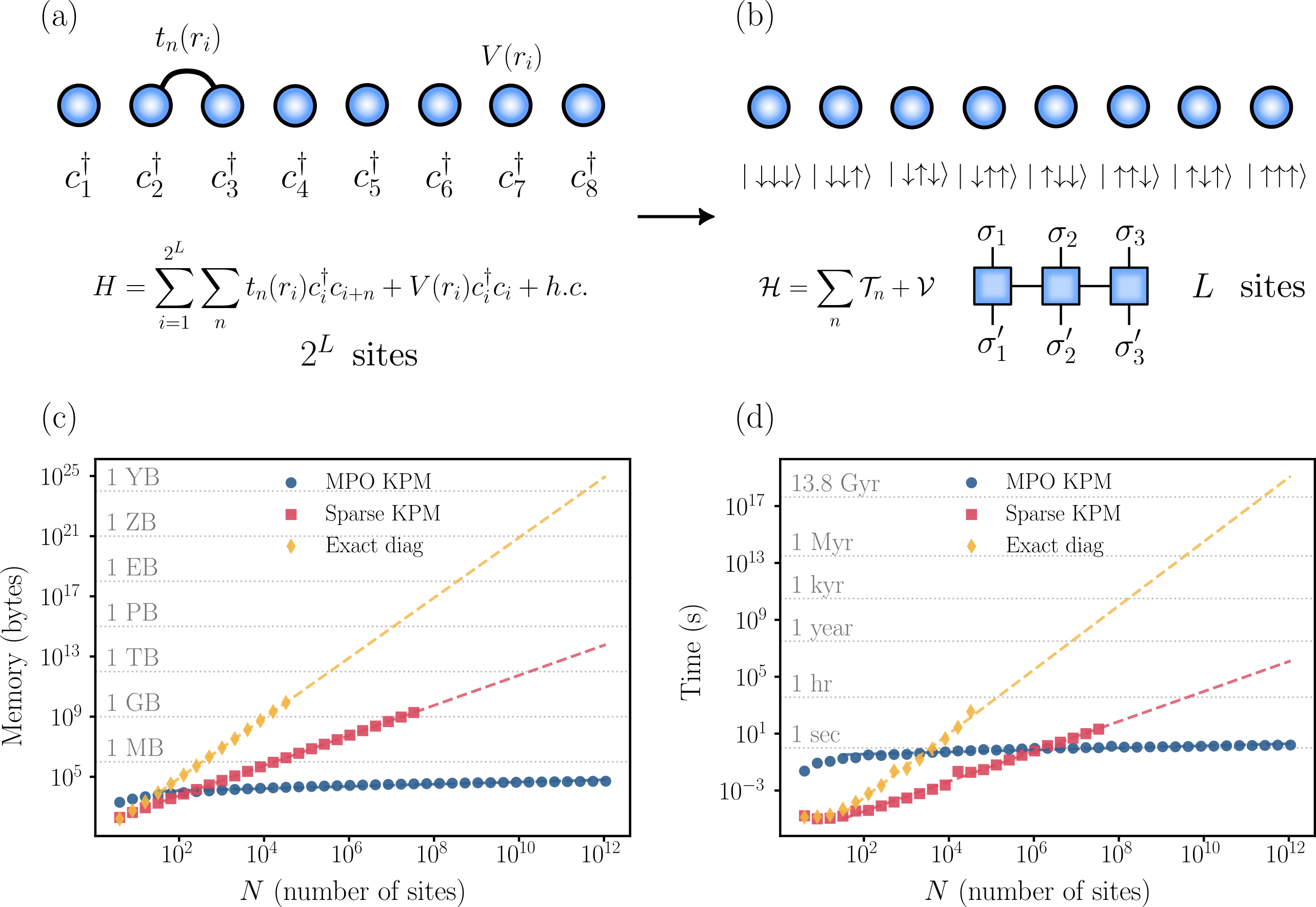}
  \caption{Mapping and benchmarks for a 1D tight-binding chain model.
    (a) Tight-binding model with $N = 2^3 = 8$ sites in real space and its
    $2^3 \times 2^3$ Hamiltonian matrix with hoppings $t_n(r_i)$ and on-site potential $V(r_i)$.
    (b) Equivalent pseudospin chain of $L=\log_2(N)=3$ sites; spin states $(\sigma_1,\sigma_2,\sigma_3)$ index real-space sites
    and the Hamiltonian is encoded as a 3-core MPO. The $n$th neighbour hoppings are encoded as MPOs $\mathcal{T}_n$ and the on-site term as an MPO $\mathcal{V}$
    (c) Memory cost of full, sparse, and MPO storage of the Hamiltonian.
    (d) Computation time for a local density of states via exact
    diagonalization, sparse-matrix KPM, and the MPO KPM implemented in
    \pkg{TensorBinding.jl}. Dahed lines are estimate, scatter explicit calculations.}
  \label{fig:mapping_benchmarks}
\end{figure}

For the sake of completeness, here we provide a brief summary of the
tensor network methodology that our algorithms exploit.
Matrix Product Operators (MPO) and Matrix Product States (MPS) are tensor-network
factorizations of quantum operators and states into one-dimensional (1D) chains of low-rank
tensors \cite{schollwock2011,orus2014,orus2019}. Each tensor in an MPS or MPO carries two types of indices: physical indices, which are used to index local degrees of freedom, and virtual or bond indices, which are contracted with its neighbors in the chain. An MPS representing a multi-index tensor $\hat{M}$ reads

\begin{equation}
\hat{M}^{\sigma_1\sigma_2\cdots\sigma_L}=\sum_{i,j,k,\dots,m,l}\Xi_{ij}^{\sigma_1}\Xi_{jk}^{\sigma_2}\cdots \Xi_{ml}^{\sigma_L}
\label{eq:generic_MPS}
\end{equation}
where $\sigma_i$ are the physical indices and $i,j,k,\dots$ are the virtual indices. Similarly, an MPO representing a multi-index tensor operator $\hat{O}$ reads

\begin{equation}
\hat{O}^{\sigma_1\sigma_2\cdots\sigma_L}_{\sigma_1'\sigma_2'\cdots\sigma_L'}=\sum_{i,j,k,\dots,m,l}\Xi_{ij}^{\sigma_1\sigma_1'}\Xi_{jk}^{\sigma_2\sigma_2'}\cdots \Xi_{ml}^{\sigma_L\sigma_L'},
\label{eq:generic_MPO}
\end{equation}
where instead of a single set of physical indices $\{\sigma_1,\dots,\sigma_L\}$ as in Eq. (\ref{eq:generic_MPS}), the MPO of Eq. (\ref{eq:generic_MPO}) now holds both unprimed and primed $\{\sigma_1',\dots,\sigma_L'\}$ physical indices, and so does each tensor core in the MPO representation.
The bond dimensions $\chi$ of an MPS or MPO correspond to the dimensions of the virtual indices and control the amount of ``complexity" \footnote{In many-body physics, the bond dimension captures a measure of the entanglement entropy of the system \cite{schollwock2011,orus2014}.} retained in the operator.
Algorithms such as iterated singular value decompositions \cite{oseledets2011,schollwock2011} or tensor cross interpolation \cite{oseledets2010,ritter2024,nunez2025} allow for a controlled decomposition or construction of MPS or MPO objects in the form of Eq. (\ref{eq:generic_MPS}) and (\ref{eq:generic_MPO}), that exactly or approximately recover generic multi-index tensors. 
In this vein, the quantics tensor train representation works by mapping real-space single-particle states $|i\rangle = |\sigma_1,\ldots,\sigma_L\rangle$, where each $\sigma_k=\uparrow,\downarrow$ constitutes a binary label $(\sigma_1\cdots\sigma_L)_2 = i$ encoding the real-space site index, to spin basis-states of a pseudo-spin Hamiltonian which can be thought of as a multi-index tensor with indices $\sigma_1\cdots\sigma_L$. The mapping between the $N$-site single-particle Hilbert space and a chain
of $L=\log_2(N)$ pseudospins is illustrated in Fig.~\ref{fig:mapping_benchmarks}.
Fig.~\ref{fig:mapping_benchmarks}(a) and Fig.~\ref{fig:mapping_benchmarks}(b)
show a minimal $N = 2^3 = 8$-site example: the
real-space chain and its $8 \times 8$ Hamiltonian matrix (a) are exactly
represented by a three-core MPO on a pseudospin chain of $L = 3$ sites 
(b). The memory footprint of the MPO Hamiltonian grows only
logarithmically with $N$, in sharp contrast to the polynomial scaling of
full and sparse-matrix storage as observed in Fig.~\ref{fig:mapping_benchmarks}(c), where notably for a $10^{12}$-site system the storage would reach Yottabyte scales of memory.
For many physically relevant tight-binding Hamiltonians in one and two spatial dimensions, this encoding
keeps $\chi \sim 10-50$ regardless of system size, making it possible to store and manipulate Hamiltonians with up to billions of sites on workstation-scale hardware.

We now explicitly describe the algorithm that allows for the construction of MPOs representing tight-binding Hamiltonians, as implemented in \pkg{TensorBinding.jl}.
The key observation is that a tight-binding Hamiltonian of the form in Eq.
\eqref{eq:tb} decomposes into two structurally distinct classes of terms, namely
on-site contributions, which are diagonal in the real-space site basis, and
hopping contributions, which connect pairs of sites.
These two classes are constructed as independent MPOs and combined by addition,
making the full Hamiltonian MPO a sum of simple building blocks each with
small bond dimension.

In the usual second quantization notation, a tight-binding Hamiltonian takes the general form
\begin{equation}
  \hat{H} = \sum_{i,j} t^{\alpha\beta}_{ij}\, c_i^\dagger c_j
    + V_i^{\alpha\beta}\, c_i^\dagger c_i + \mathrm{h.c.},
  \label{eq:tb}
\end{equation}
where $i,j$ label lattice sites and $\alpha,\beta$ denote internal degrees of freedom such as sublattice, spin, orbital, or Nambu indices. The hopping amplitudes $t^{\alpha\beta}_{ij}$ encode couplings between sites, while $V^{\alpha\beta}_{i}$ is an on-site potential that may be non-diagonal over the internal degrees of freedom.

Upon mapping to the pseudo-spin system, on-site potentials are represented as operators $\hat{V}$ which take the form of diagonal MPOs.
In the quantics encoding the site index $i$ is written in binary, so the
tensor network representation of a compressible
on-site function $V(i)$ can be obtained with QTCI, yielding
a diagonal MPO whose maxiumum bond dimension is set by the smoothness of $V$ and is nearly independent
of the system size $N$.

On the other hand, an $n$th-nearest-neighbour hopping on a chain of $N = 2^L$ sites requires an operator that maps $|i\rangle \mapsto |i+n\rangle$. The central building block for this construction is a compact MPO we refer to as the shift tensor, $\hat{S}^{(n)}$, which implements the required shift in the pseudospin representation. Several constructions for this object are possible, ranging from powers of an ``addition by 1" operator as described in \cite{sun2025,antao2026} or the so-called magic tensor, which directly realizes an ``addition by $n$" operator as described in \cite{waintal2026}.
Its construction amounts to implementing binary addition when shifting $i$ by $n$, propagating carries from the least
significant to the most significant pseudospin.
Each carried step is local in the tensor chain and involves only Pauli raising
and lowering operators $\hat{\sigma}^\pm$ on neighboring pseudospin, so $\hat{S}^{(n)}$ has
bond dimension $\chi$ independent of $N$. An illustration of this operator's construction for a three-pseudospin system is detailed in Appendix~\ref{app:shift} .
The complete $n$th-nearest-neighbor hopping MPO is then assembled as
\begin{equation}
  \hat{H}_{\rm hop}^{(n)} = \hat{\mathcal{T}}_n\,\hat{S}^{(n)} +\bigl[\hat{S}^{(n)}\bigr]^\dagger\hat{\mathcal{T}}_n^\dagger,
\end{equation}
where $[\hat{S}^{(n)}]^\dagger = \hat{S}^{(-n)}$ shifts in the opposite direction.
Here, spatially modulated amplitudes $t_{i,i+n}(i)$ are incorporated by taking the product of the shift MPO with the diagonal MPO which encodes $t_{i,i+n}(i)$, represented by $\hat{\mathcal{T}}_n$. This approach keeps the construction entirely within MPO algebra since the product of $\hat{S}^{(n)}$ with $\hat{\mathcal{T}}_n$ represents the full hopping operator. 

In higher dimensions, the construction is analogous. For instance, a rectangular lattice of $2^{L_x} \times 2^{L_y}$ sites is mapped onto a
1D chain of $L = L_x + L_y$ pseudospins via the row-major encoding
\begin{equation}
  n(i_x, i_y) = i_x + i_y\,2^{L_x}.
  \label{eq:rowmajor}
\end{equation}
Under this mapping a lattice displacement $(\Delta_x, \Delta_y)$ becomes a shift
$q = \Delta_x + \Delta_y 2^{L_x}$ in the effective 1D chain, so the same shift-tensor
machinery applies without modification. 
The only subtlety is that the periodic boundary of the 1D chain would otherwise
connect sites at the end of one row to sites at the beginning of the next.
These spurious wrap-arounds can be suppressed by masking MPOs that project out
site pairs straddling a row boundary, restoring the intended open or periodic
boundary conditions, as described in full in Appendix~\ref{app:2d}.
Any Bravais lattice geometry, whether square, triangular, honeycomb, kagom\'e, and beyond, can then be encoded by enumerating the set of physical displacement vectors
$\{(\Delta_{x_\alpha}, \Delta_{y_\alpha)}\}$  together with their amplitudes, which may be
uniform, direction-dependent, sublattice-resolved, or an arbitrary function of
position further compressed via QTCI.

The modularity of this construction makes it straightforward to generalize: higher-dimensional lattices are handled by extending the row-major encoding
to additional index blocks ($n = i_x + 2^{L_x}i_y + 2^{L_x+L_y}i_z + \cdots$), and arbitrary hopping ranges are
selected simply by choosing the corresponding shift distance $q$.
The result is a unified, dimension-agnostic MPO-based framework applicable to
the full breadth of tight-binding models encountered in practice.

\subsection{Implementation details in \pkg{TensorBinding.jl}}
In the \pkg{TensorBinding.jl} framework, 
several pre-built Hamiltonians can be called from a named geometry string covering common 1D and
2D lattices, namely 1D chains, SSH models \cite{su1979}, Aubry-Andr\'e-Harper quasicrystals \cite{aubry1980,hofstadter1976},
and 2D square, triangular, honeycomb, kagom\'e, Lieb, and dice lattices. Sublattice-aware models carry an explicit sublattice index in the tensor
train, enabling the calculation of sublattice-resolved observables.
Arbitrary $n$th-nearest-neighbor hoppings can be added dynamically, with
amplitudes modulated uniformly, by bond direction, by site, or by full
position-and-direction dependence, making it straightforward to encode complex
phase modulations such as Haldane NNN hoppings \cite{haldane1988} without
modifying the underlying MPO object.
Hamiltonian objects can also be constructed from externally provided MPOs and
geometry functions.

Many problems of current interest require degrees of freedom beyond a scalar
hopping amplitude: spin-orbit coupling in topological insulators \cite{hasan2010},
Nambu structure in superconducting systems \cite{kitaev2001,alicea2012,cao2018sc,lu2019,park2021},
or the layer index in moir\'e heterostructures \cite{cao2018sc,tran2019,andrei2020,andrei2021}.
The MPO representation accommodates these through composable mutating
operations that append spin-$\tfrac{1}{2}$, Nambu (BdG), layer, sublattice,
or T-junction branch indices to an existing Hamiltonian, with the
auxiliary-index structure tracked throughout, following the construction detailed in Appendix~\ref{app:aux}.
For instance, T-junctions, where three 1D chains meet at a central region, are supported via a dedicated dim-3 branch index, which enables
the study of multi-terminal geometries relevant to, e.g., Majorana braiding
protocols and multi-wire topological networks \cite{kitaev2001,alicea2012}.
Another example pertains to twisted multilayers with exponentially decaying interlayer coupling, which are
supported via dedicated QTCI-compressed constructors, enabling moir\'e
systems to be built at scales inaccessible to explicit matrix methods. 

\section{Single Particle Physics}\label{sec:single_particle}

\subsection{Real- and momentum-space spectral functions}

Accessing the local density of states and momentum-space spectral functions of large systems
without diagonalization is essential for moir\'e and quasiperiodic materials,
where the spectrum is dense and diagonalization-based approaches are
intractable \cite{joao2020,yuan2022,andelkovic2018,gimenez2023,kuang2025,quantica}.
We employ the Kernel Polynomial Method (KPM) \cite{weisse2006,holzner2011}, which expands spectral quantities in Chebyshev polynomials of the Hamiltonian
evaluated iteratively via $T_n(H) = 2H T_{n-1}(H) - T_{n-2}(H)$, and with $T_0=\mathbbm{1}$, $T_1=H$, entirely in MPO arithmetic.
KPM techniques have been widely used with sparse tight-binding Hamiltonians \cite{fan2021,mucciolo2010,covaci2010,garcia2015,andelkovic2018,joao2020}, with the drawback that, in its sparse representation, KPM only allows access to properties of a single vector at a time. This stems from the fact that KPM algorithms are applied at the vector level as $|v \rangle = T_n(H) |\alpha\rangle $, with $|\alpha \rangle$ a specific initial vector, whereas performing KPM directly on the Hamiltonian itself as $T_n(H)$ is prohibitively large due to memory constraints.
In stark contrast, tensor networks enable applying KPM directly to the tensor network Hamiltonian $T_n(H)$, computing the Chebyshev polynomial of the Hamiltonian itself explicitly, and accessing spectral information of the entire system at once via tensor network contraction.
Summations over these polynomials can be used to approximate functions over the interval $(-1,1)$, and in particular, after rescaling a tight-binding Hamiltonian's spectrum into this range, one may approximate spectral functions such as

\begin{equation}
\hat{A}(\omega)\equiv\delta(\omega-\hat{H})\approx\sum_{n=0}^{N_c} g_n \hat{\mu}_n\frac{2-\delta_{n,0}}{\pi \sqrt{1-\omega^{2}}}T_{n}(\omega), 
\label{eq:Aw_Chebyshev}
\end{equation}
where the weights of the expansion are themselves Chebyshev polynomials of the Hamiltonian MPO  $\hat{\mu}_n=T_n(\hat{H})$ and where the frequency polynomials can be directly expressed as $T_n(\omega)=\cos(n\arccos(\omega))$. Additional kernels $g_n$ can be used under specific conditions for the smoothening or speed-up of convergence of the KPM expansion \cite{weisse2006,jackson1912,yi2025}.  It is worth noting that the rescaling of the Hamiltonian necessary for the KPM algorithm may be performed efficiently, even if its spectral support is unknown beforehand, by leveraging the DMRG algorithm \cite{white1992}. One can find the highest and lowest energy eigenstates and eigenvalues of the Hamiltonian and compute the scale and center of its energy spectrum. This is implemented directly in \pkg{TensorBinding.jl}. Similar Chebyshev expansions can be performed both for the density matrix and Green's function by changing the frequency-dependent coefficients of Eq. (\ref{eq:Aw_Chebyshev}) accordingly \cite{weisse2006,holzner2011}.

Figure \ref{fig:mapping_benchmarks}(d) quantifies the computational advantage over conventional
methods for a simple system consisting of a uniform 1D tight-binding chain: the MPO KPM implemented in \pkg{TensorBinding.jl} reaches system
sizes well beyond those accessible to exact diagonalization or even sparse-matrix
KPM. As a reference, beyond memory constraints,
solving system sizes of $10^{12}$ on a single core with exact diagonalization 
would take one billion years, whereas our tensor network algorithm
performs the calculation in a few seconds. In addition to the previously discussed memory requirements, these constraints render conventional methods unrealistic for ultra-large length scales.

Within the \pkg{TensorBinding.jl} framework, three strategies address different physical regimes.
In ``MPO mode", each $T_n(H)$ is stored as an MPO and the full Chebyshev list is
contracted against a position projector, giving spatially-resolved local
quantities.
In ``diagonal mode", only the diagonal of each $T_n(\hat{H})$ is extracted in a single
pass at a cost of $\mathcal{O}(N_{\rm cheb})$ MPO-MPS products and no
intermediate storage.
In ``MPS mode", a reference state $|\phi_n\rangle = T_n(H)|\psi_0\rangle$ is
propagated instead, and this is the natural route for specific regimes
where bond-dimension growth may otherwise become unmanageable, which, as we elaborate below, is relevant to excitonic calculations.
This strategy is analogous to the KPM vector algorithm used with sparse Hamiltonians, but expressed in MPS form. The tensor-network structure of all three pathways is detailed in Appendix~\ref{app:KPM} .

The momentum-resolved spectral function $A(\mathbf{k},\omega)$, which enables one to
reconstruct band structures, is
obtained by combining the KPM operator with the Quantum Fourier Transform
MPO $\hat{U}_{\rm QFT}$ native to the pseudospin encoding \cite{khoromskij2011, chen2023fourier, moustaj2025momentum, niedermeier2024}. This operator encodes the transformation as

\begin{equation}
   \hat{U}_\text{QFT} |k\rangle =\frac{1}{\sqrt{N}}\bigotimes_{\ell=1}^L\sum_{\sigma_\ell=0}^1e^{2\pi ik\sigma_\ell2^{-\ell}}|\sigma_\ell\rangle
            = \bigotimes_{\ell=1}^L\frac{1}{\sqrt{2}}\left[|\!\downarrow\rangle+ e^{2\pi ik2^{-\ell}}|\!\uparrow\rangle\right]
    \label{eq:Fourier_mpo}
\end{equation}
Specifically, taking $\hat{A}(\omega)$ as the real-space spectral MPO,
the $k$-space counterpart is $\hat{\Tilde{A}}(\omega) = \hat{U}_{\rm QFT}\,\hat{A}(\omega)\,\hat{U}_{\rm QFT}^\dagger,$
where the operator $\hat{\Tilde{A}}(\omega)$ remains an MPO
throughout the computation, as detailed in Appendix \ref{app:momentum}, so that the spectral function can be directly extracted by contracting with localized momentum-space MPSs $|\mathbf{k}\rangle$ as
$A(\mathbf{k},\omega) = \langle\mathbf{k}|\hat{\Tilde{A}}(\omega)|\mathbf{k}\rangle$.
This gives direct access to band dispersion at system sizes where full
unfolding onto the moir\'e Brillouin zone would be prohibitive. Additionally, using projection operators onto specific regions in real space \cite{moustaj2025momentum}, as elaborated on in Appendix \ref{app:momentum}, this allows
probing local band structures of moir\'e modulated systems  
of high relevance to model quantum twisting microscope experiments \cite{inbar2023,wei2025,birkbeck2025,xiao2026}.

\subsection{Real-space topological invariants}

Topological invariants such as the Chern number are conventionally computed
in reciprocal space \cite{thouless1982}, but quasicrystals and disordered or inhomogeneous
systems lack Bloch states entirely, requiring a purely real-space formulation.
The real-space approach is built on the single-particle density matrix,
which at zero temperature coincides with the ground-state projector,
\begin{equation}
  \hat{P}\equiv\hat{\rho} = \Theta(H-\epsilon_F)= \sum_{n \in \mathrm{occ}} |\psi_n\rangle\langle\psi_n|,
\end{equation}
where $\epsilon_F$ is the Fermi energy. From $\hat{P}$, topological invariants can be extracted without diagonalization
\cite{bianco2011,chen2023marker,sykes2021,sykes2022,PhysRevB.109.014206,han2023,sahlberg2023,duncan2020,manna2024}.
Closely related real-space constructions extend this projector-based strategy to the local quantum geometry and fidelity markers of inhomogeneous and disordered phases \cite{peotta2015,torma2022,martinezromeral2025,desousa2023,oliveira2024}.
Three strategies to build $\hat{\rho}$ within MPO algebra are available within \pkg{TensorBinding.jl}: KPM-based Chebyshev
expansion, McWeeny purification \cite{mcweeny1960}, and SP2 purification \cite{niklasson2002}. Purification methods construct $\hat{\rho}$ by recognizing that the ground-state projector is the Heaviside step function of the Hamiltonian, $\hat{P} = \Theta(\hat{H}-\varepsilon_F)$, and that this step function, with eigenvalues exactly 0 or 1, is the unique stable fixed point of certain polynomial maps. Starting from an initial approximation with spectrum in $[0,1]$, one iterates such a map until convergence. McWeeny purification uses $\hat{\rho}_{n+1} = 3\hat{\rho}_n^2 - 2\hat{\rho}_n^3$
which converges directly to the fixed point, while SP2 purification \cite{niklasson2002} alternates between two quadratic branches, $\hat{\rho}_{n+1} = \hat{\rho}_n^2$ if  $\text{tr}(\hat{\rho}_n^2) \geq N_e$, or $\hat{\rho}_{n+1}= 2\hat{\rho}_n - \hat{\rho}_n^2$ if $\text{tr}(\hat{\rho}_n^2) < N_e $
with the branch chosen at each step to keep the trace equal to the target electron count $N_e$. Together with a KPM-based Chebyshev expansion of $\Theta(\mu-\hat{H})$, all three algorithms converge to the same projector, however the two purification approaches provide significantly improved accuracy and speed over a pure KPM method to compute the projector \cite{sun2025, antao2026} in most cases \cite{sobrosa2026}

Using the ground-state projector, the 2D Chern marker \cite{bianco2011,chen2023marker} can then be computed as
\begin{equation}
  C(\mathbf{r}) = 4\pi i\,
  \langle\mathbf{r}|\hat{Q}\hat{x}\hat{P}\hat{y}\hat{Q} - \hat{P}\hat{x}\hat{Q}\hat{y}\hat{P}|\mathbf{r}\rangle,
\end{equation}
and the 1D winding-number density \cite{chen2023marker} as
\begin{equation}
  W(\mathbf{r}) =
  \langle\mathbf{r}|\,\hat{\sigma}_z\,(\hat{P}\hat{x}\hat{Q} + \hat{Q}\hat{x}\hat{P})\,|\mathbf{r}\rangle.
\end{equation}
These give spatially resolved topological information at every unit cell, and depend on the projectors $\hat{P}$ and
$\hat{Q} = 1 - \hat{P}$, as well as position operators $\hat{x}$, $\hat{y}$.
Evaluating these markers requires regularizing the position operators
to keep the MPO bond dimension compact. This procedure is described in detail in Appendix~\ref{app:topology} .
This methodology allows mapping the real-space topology of quasicrystalline Chern mosaics at scales of hundreds of millions of sites \cite{antao2026}, where the local Chern marker reveals coexisting topological phases inaccessible to reciprocal-space methods.

Additionally, valley projected Chern markers \cite{xiao2007,zhang2013} can be accessed using the real-space formulation of the valley projectors. This allows for the characterization of valley topology useful for instance in periodically buckled or sublattice imbalanced graphene \cite{manesco2021}.

\begin{figure}[!t]
    \centering
    \includegraphics[]{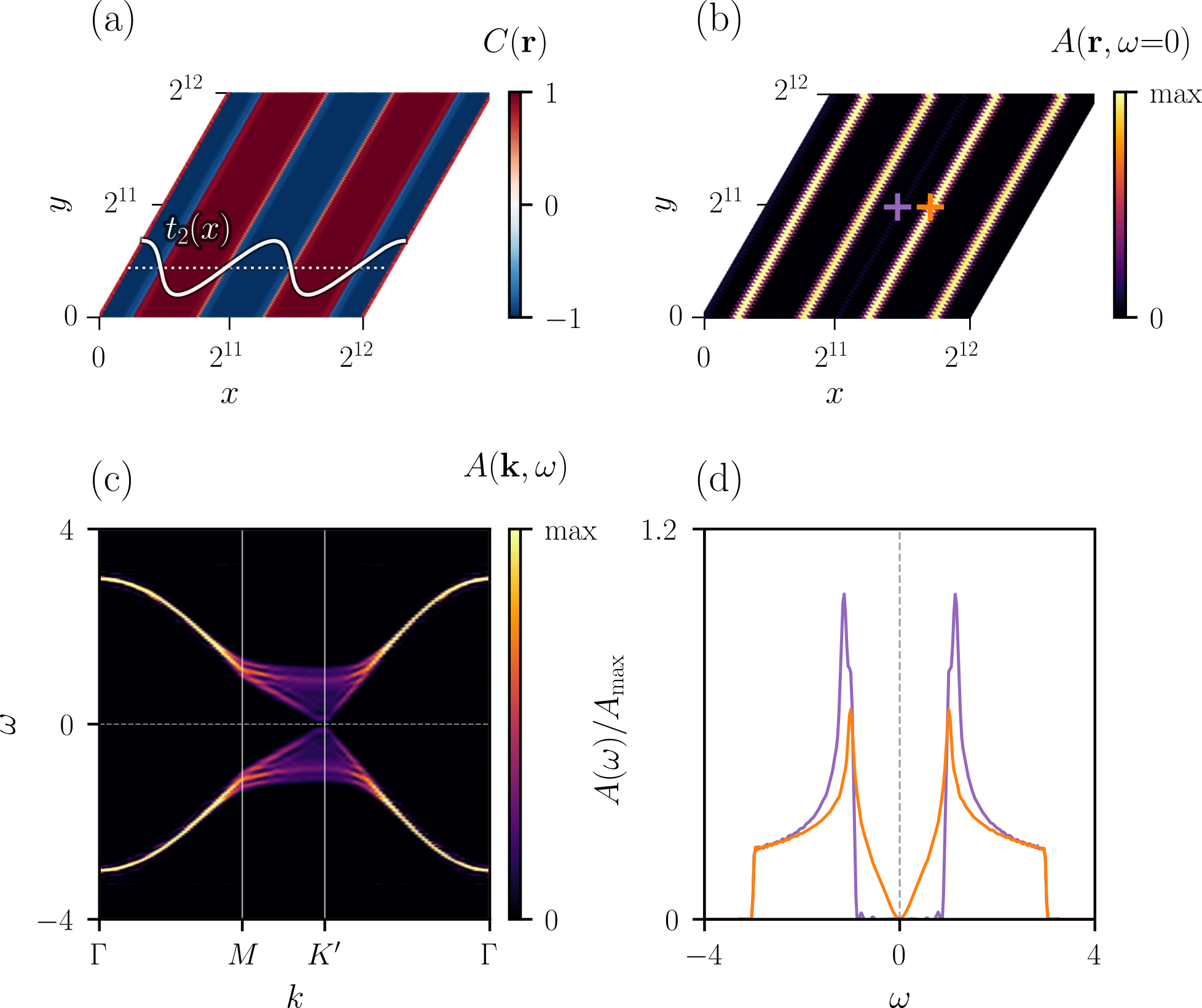}
    \caption{Single-particle observables for the spatially modulated Haldane model of Eq.~\eqref{eq:haldane_mod} with $t_2(x) = \bar{t}_2\cos(2\pi x / \lambda)$.
(a) Real-space Chern marker, with well-defined integer stripes reflecting alternating topological and trivial phases of period $\lambda=N_x/2$.
(b) Zero-frequency LDOS, with intensity peaks localised at the domain walls between topological phases.
(c) Band structure showing mini-band folding at the superlattice wavevector and a weak imprint of the gap-traversing chiral edge states.
(d) Frequency-resolved LDOS in the bulk (top) and at a domain-wall edge (bottom), showing the gapped bulk spectrum and the in-gap chiral edge branch.
The parameters used for these calculations are $t_1=1$, $\bar{t}_2=0.3t_1$, $\phi=\pi/2$, $M=0$, and $\lambda=N_x/2$, where $N_x=N_y=2^{12}$ is the number of unit cells in the $x$ and $y$ directions.}
    
    \label{fig:single_particle_panel}
\end{figure}

As a concrete example for all the single-particle methodologies described above, we take the Haldane model \cite{haldane1988} on a
honeycomb lattice with a spatially modulated next-nearest-neighbor amplitude.
The Hamiltonian is
\begin{equation}
  \hat{H} = t_1 \sum_{\langle i,j\rangle} c_i^\dagger c_j
    + \sum_{\langle\langle i,j\rangle\rangle}
        t_2(x_i)\,e^{i\phi_{ij}}\,c_i^\dagger c_j
    + M\sum_i \xi_i\,c_i^\dagger c_i + \mathrm{h.c.},
  \label{eq:haldane_mod}
\end{equation}
where $t_1$ is the nearest-neighbor hopping, $\phi_{ij} =\pm\phi$ is the
Haldane flux (sign fixed by bond orientation), $M$ is a sublattice mass, and
$\xi_i = \pm 1$ on the two sublattices.
The next-nearest-neighbour amplitude is modulated along $x$ as $t_2(x) = \bar{t}_2\cos(2\pi x / \lambda)$,
with mean amplitude $\bar{t}_2$, and period $\lambda$. 
This periodic variation drives the system through alternating topological and
trivial regions, producing a Chern mosaic whose domain structure is set by
$\lambda$.
Fig.~\ref{fig:single_particle_panel} shows the results of a full single-particle
analysis within \pkg{TensorBinding.jl}. The real-space Chern marker (panel a) takes well-defined integer values in
stripes of width $\sim\lambda/2$, directly resolving the alternating
$C = \pm 1$.
The zero-frequency local density of states Fig.~\ref{fig:single_particle_panel}(b) peaks at the boundaries
between topological and trivial stripes, where chiral edge states localize at
each internal domain wall.
The band structure Fig.~\ref{fig:single_particle_panel}(c) retains the Dirac-cone topology of the uniform
Haldane model but acquires mini-band folding at wavevector $2\pi/\lambda$,
reflecting the superlattice periodicity. The frequency-resolved LDOS Fig.~\ref{fig:single_particle_panel}(d), evaluated separately in the bulk and
at a domain-wall edge, shows the gapped bulk spectrum together with the
in-gap edge branch that disperses across the bulk gap, confirming the
topological origin of the boundary modes.

\subsection{Non-Hermitian systems}

Non-Hermitian tight-binding models naturally describe open systems with gain,
loss, and non-reciprocal hopping, including photonic lattices, or dissipative
electronic systems\cite{ashida2020,bergholtz2021,gong2018,helbig2020,kunst2018,okuma2020,yao2018,song2019}.
In the package, spatially dependent gain or
loss is represented by an imaginary diagonal MPO, while asymmetric forward and
backward hopping amplitudes generate non-reciprocal Hamiltonians.
To evaluate spectral observables, a non-Hermitian Hamiltonian is mapped to an
energy-dependent Hermitian auxiliary problem\cite{chen2023nh}. For a complex energy
$z=E+i\eta$, we define
\begin{equation}
  \hat{\mathcal{H}}(z) =
  \begin{pmatrix}
    0 & z\hat{I} - \hat{H} \\
    z^*\hat{I} - \hat{H}^\dagger & 0
  \end{pmatrix},
  \label{eq:nh_hermitization}
\end{equation}
implemented by adding a two-dimensional auxiliary block index to the MPO.
The
resulting Hermitian Hamiltonian can be treated with a modified KPM algorithm \cite{chen2023nh,chen2024nh,sun2026nonhermitian}, avoiding explicit
diagonalization of the non-Hermitian matrix, with the modified recursion derived in Appendix~\ref{app:nhkpm}.
This construction gives access to the non-Hermitian spectral response and
spatially resolved LDOS at fixed complex energy. The default implementation
reconstructs the local spectral profile as an MPS to control bond-dimension
growth, while a higher-cost all-site MPO reconstruction is also available when
one wants the full LDOS object in a single calculation\cite{chen2023nh,chen2024nh,sun2026nonhermitian}.

\begin{figure}[!t]
    \centering
    \includegraphics{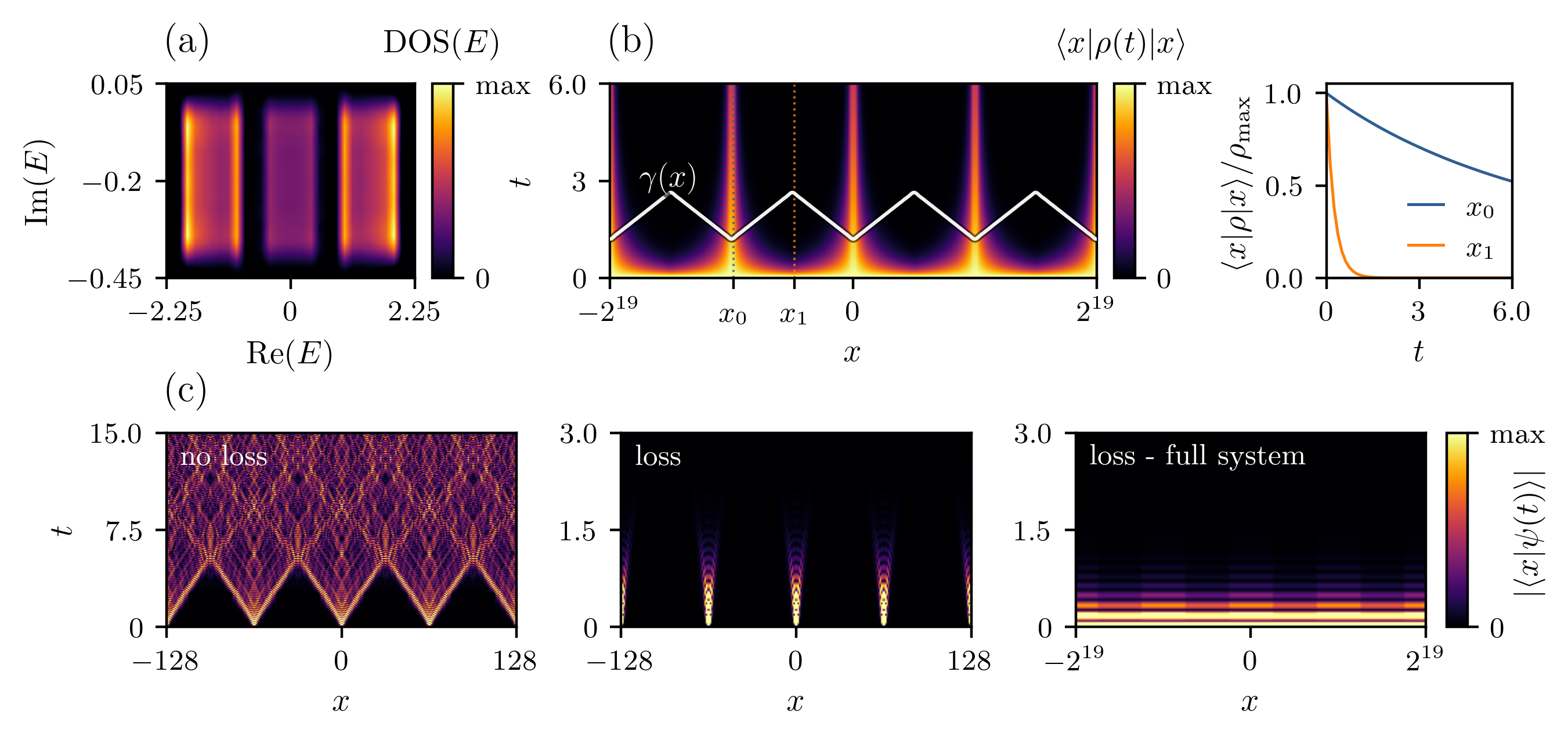}
    \caption{Non-Hermitian observables for a 1D AAH chain with the modulated loss
    $\gamma(x)$ following a triangular pattern under open boundary conditions.
    (a) Spectral weight $A(\mathrm{Re}\,z, \mathrm{Im}\,z)$ in the complex
    energy plane from the NH KPM algorithm.
    (b) Space-time map of the diagonal density $\langle x|\rho(t)|x\rangle$
    evolved from the half-filled ground state via the non-Hermitian
    von~Neumann equation (RK4),
    and site-density traces at the loss minimum ($x_0=N/4$,
    $\gamma = 0$) and loss maximum ($x = 3N/8$, $\gamma = 2\gamma_0$).
    (c) Single-particle amplitude $|\langle x|\psi(t)\rangle|$
    from TDVP for a comb wavepacket shown in the central region of a Hermitian chain (left) and a lossy chain (middle), zoomed in
    a region of width $w=256$ sites, and across the full spatial extent of the lossy chain (right). The length of the chain in (a,b,c) is $2^20$.
    Parameters: $t = 1$, $V_{\rm{AAH}}=0.5t$, and $\gamma_0 = 0.2t$.}
    \label{fig:nh_panel}
\end{figure}

As a concrete example, we consider an Aubry-Andre-Harper chain \cite{aubry1980} with $N = 2^{20}$ sites under
open boundary conditions, 
\begin{equation}
    \hat{H}_0=t\sum_{n=1}^{N} c_n^\dagger c_{n+1} + \sum_{n=1}^{N} V_{\rm{AAH}}\cos\left(2\pi x_n/\phi\right)c_n^\dagger c_n +\rm{h.c.},
\end{equation}
where $\phi=(1+\sqrt{5})/2$ is the golden ratio, and an imaginary on-site
potential $\hat{\Gamma}=\sum_n \gamma(x_n)c_n^\dagger c_n,$ with $\gamma(x_n)$ following a triangular wave modulation pattern with a large-scale wavelength of $N/4$ and modulation strength $\gamma_0$.
The resulting non-Hermitian Hamiltonian $\hat{H} = \hat{H}_0 - i\hat{\Gamma}$ has a loss-free pocket at
$x_0=N/4$ and a maximum loss one $2\gamma_0$ at $x_1 = 3N/8$.
Figure~\ref{fig:nh_panel} shows the observables computed within
\pkg{TensorBinding.jl} for this model.
Panel (a) shows the spectral weight across the complex energy plane from the NH KPM algorithm: states populate the lower half-plane, with imaginary
parts encoding mode-dependent decay rates.

\subsection{Out-of-equilibrium dynamics}
Another core functionality of the \pkg{TensorBinding.jl} framework pertains to the evaluation of out-of-equilibrium time-dependent systems. 
Driven quantum systems such as Floquet materials, pump-probe experiments, or nonlinear
optical responses require real-time propagation of large lattice models under
time-dependent fields, a regime where exact diagonalization is doubly
intractable \cite{oka2019,mciver2020,rudner2020,delatorre2021,huber2026,cai2026,holtzmann2026}. 
Two complementary routes are available within the MPO/MPS framework. Firstly, the Time-Dependent Variational Principle (TDVP) \cite{haegeman2011,haegeman2016,paeckel2019} integrates the
Schr\"{o}dinger equation directly on the MPS manifold, keeping the bond
dimension bounded at every step.
Both static and explicitly time-dependent Hamiltonians $\hat{H}(t)$ are supported,
with $\hat{H}$ evaluated at the midpoint of each time step. Secondly, for open or mixed-state systems the single-particle density matrix $\hat{\rho}(t)$
is evolved under the von~Neumann equation $\dot{\hat{\rho}} = -i[\hat{H}(t),\hat{\rho}]$ via a fourth-order Runge-Kutta scheme entirely in MPO arithmetic, as detailed in Appendix~\ref{app:rk4}. This allows for the computation of, for instance, time-dependent currents or particle densities, as well as high-harmonic-generation quantities from the evolution of time-dependent Hamiltonians with hoppings modulated by time-dependent Peierls phases.

Here, we consider another example in which the time evolution of large-scale systems may be interesting, namely, the time evolution of the non-Hermitian Hamiltonian presented in Sec.~\ref{sec:single_particle}C. 
Figure~\ref{fig:nh_panel}(b) combines the space-time map of the many-body diagonal density
$\langle x|\rho(t)|x\rangle$, obtained by propagating the half-filled
ground-state density matrix under the non-Hermitian von~Neumann equation via
the RK4 solver, with site-density traces at the loss minimum
$x_0$ and and loss maximum $x_1$: particle weight accumulates progressively
at the loss-free pockets while the high-loss sites decay.
Panel (d) compares single-particle wavepacket dynamics via TDVP for two
cases: the Hermitian chain and a comb wavepacket initialised at sites separated by strides of length $\ell=64$. The density $|\langle x|\psi(t)\rangle|$ demonstrates
that attenuation is strongly position-dependent. The panel also showcases the multiscale capabilities of \pkg{TensorBinding.jl}, as the last figure depicts the evolution of the wave packet across the full system width.

\section{Correlated Phenomena}\label{sec:correlated}
Electron-electron interactions represent the driving force behind a variety of phenomena in quantum materials \cite{keimer2017}, including symmetry-broken magnetic and charge ordering \cite{imada1998,cao2018insulator}, unconventional superconductivity \cite{scalapino2012,cao2018sc}, and excitonic bound states \cite{tran2019,wu2018}, among others, none of which are captured by the single-particle framework alone. Many-body extensions implemented in \pkg{TensorBinding.jl} address this within the MPO framework.

\subsection{Self-consistent field calculation}
Correlation-driven instabilities, such as symmetry-broken order in moir\'e systems, can be studied via a Hartree-Fock-type self-consistent field loop.
Starting from a general density-density two-body interaction
\begin{equation}
  \hat{H}_{\rm int} = \tfrac{1}{2}\sum_{i,j,s,s'}
    V_{ij}^{ss'}\,\hat{n}_{is}\hat{n}_{js'},
\end{equation}
where $s=\Uparrow,\Downarrow$ corresponds to the physical electron spin, a Wick decoupling replaces each pair of number operators by their mean-field
expectation value, on top of which are considered single-particle fluctuations, yielding a self-consistent single-particle
Hamiltonian
\begin{equation}
  \hat{H}_{\rm MF} = \sum_{i,j,s}
    \bigl[t_{ij,s} + \Sigma^{\rm H}_{ij,s}
                        + \Sigma^{\rm F}_{ij,s}\bigr]
    c^\dagger_{is}c^{\phantom\dagger}_{js},
\end{equation}
with Hartree (direct) $\Sigma^{\rm H}_{ij,s} = \delta_{ij}\sum_{k,s'} V_{ik}^{ss'}\hat{\rho}_{kk,s'}$, and Fock (exchange) $\Sigma^{\rm F}_{ij,s} = -V_{ij}^{ss}\,\rho_{ij,s}$ self-energies evaluated from the single-particle density matrix
$\rho_{ij,s} = \langle c^\dagger_{js}c^{\phantom\dagger}_{is}\rangle$
and updated at each iteration until convergence. The construction of both self-energies as MPOs described in Appendix~\ref{app:hf}.
The SCF loop in \pkg{TensorBinding.jl} can be performed through both these Hartree and Fock channels depending on the form of $V_{ij}^{ss'}$. Aditional particular cases are optimized for in \pkg{TensorBinding.jl}, including charge-density-wave order driven by
local or long-range Coulomb repulsion, spin-singlet $s$-wave and equal-spin
$p$-wave superconducting pairing via a Bogoliubov-de Gennes loop, as well as
a general Fock exchange builder.
Here we detail the magnetic channel as a representative example.

We focus on the on-site Hubbard model \cite{hubbard1963}, with Hamiltonian
$V_{ij}^{ss'} = U_i\,\delta_{ij}(1-\delta_{ss'})$
(repulsion only between opposite spins on the same site). The Fock term
vanishes under a collinear magnetic ansatz and the decoupling reduces to
$U\hat{n}_{i\Uparrow}\hat{n}_{i\Downarrow} \approx
U(\langle\hat{n}_{i\Uparrow}\rangle\hat{n}_{i\Downarrow}
 + \langle\hat{n}_{i\Downarrow}\rangle\hat{n}_{i\Uparrow})$,
giving
\begin{equation}
  \hat{H}_{\rm MF} = \sum_{i,j,s} t_{ij}\,c^\dagger_{is}c^{\phantom\dagger}_{js}
    + \sum_{i}\bigl[(V_i + U_i\langle\hat{n}_{i\Uparrow}\rangle)\,\hat{n}_{i\Downarrow}
    + (V_i + U_i\langle\hat{n}_{i\Downarrow}\rangle)\,\hat{n}_{i\Uparrow}\bigr],
\end{equation}
where spin-up and spin-down density matrices are iterated to convergence with
local densities evaluated at each step via KPM or purification.
This particular channel has been used to compute correlated phases of super-moir\'e systems
beyond one billion sites \cite{sun2025}.
As a concrete illustration, Fig.~\ref{fig:correlated_panel} shows results for
a two-dimensional (2D) square lattice with $L_x = L_y = 12$ ($2^{24}$ sites),
hopping modulation  $t(x) = t_0 + t_{\rm amp}\cos(8\pi x/N_x)$, with $t_{\rm amp}=0.5t_0$ and Hubbard repulsion
$U = 5.5t_0$.
Fig.~\ref{fig:correlated_panel}(a) shows the converged real-space magnetization $|m_i|$: the system
develops stripe order aligned with the hopping modulation, with the strongest
magnetic moments concentrated in the regions of enhanced hopping.
Panel (b) shows the spin-summed spectral function $A(k,\omega)$ along the
high-symmetry path, computed via KPM from the converged magnetic density matrix;
the quasiparticle bands of the ordered state are clearly resolved together with
the correlation-driven gap at the Fermi energy.

\subsection{Two-body excitonic correlated states}
The second many-body extension of \pkg{TensorBinding.jl} concerns the formation of two-particle bound states, 
enabling computing excitonic physics. In moir\'e bilayers, excitons are bound electron-hole pairs whose binding
energy and spatial structure are set by competition between the single-particle
band geometry and the Coulomb interaction, requiring explicit treatment of both
degrees of freedom simultaneously \cite{tran2019,wu2018,herrera2025}.
Excitons are two-particle bound states, and span the tensor product of electron and hole
Hilbert spaces. From a tensor network perspective, the excitonic space is encoded in an interleaved two-particle MPO on $2L$ pseudospin sites, with odd (even) sites carrying the electron (hole) position pseudospins.
The electron-hole interaction $U(\mathbf{r}_e, \mathbf{r}_h)$ is compressed with QTCI and added to the single-particle kinetic terms \cite{salpeter1951,onida2002,ljungberg2015,blase2020,mattis1986}. 
The exciton spectral function is then obtained via the KPM MPS pathway,
giving access to bound states and continua in systems with more than one
billion electron-hole pairs \cite{moustaj2026excitons}. 
The Bethe-Salpeter Hamiltonian takes the form  $\hat{H}_X =\hat{T}_c\otimes\mathbbm{1} - \mathbbm{1}\otimes\hat{T}_v - \hat{U}$,
where $\hat{T}_c$ contains the single-particle conduction band terms, $\hat{T}_v$ is the contribution arising from the valence band, and $\hat{U}$ is the interaction coupling electrons and holes. For example, a Hubbard-like on-site interaction will be given by $U_{klmn}=U_k\delta_{kl}\delta_{mn}\delta_{ln}$ \cite{moustaj2026excitons}. The construction of this interleaved two-particle Hamiltonian is detailed in Appendix~\ref{app:bse}.

Fig.~\ref{fig:correlated_panel}(c) shows the exciton local density of states
$\rho(x,\omega)$ for a 1D system of $2^{20}$ electron-hole pairs ($L = 20$)
with an incommensurate on-site modulation $V(x) = V_0\cos(2\pi x/\lambda)$,
$V_0 = 1.5|t|$, and Hubbard interaction $U = 6|t|$.
Bound-state spectral weight appears below the scattering continuum and is
periodically modulated in space; the incommensurability of $V(x)$ gives rise
to moir\'e minibands visible as periodic replicas in the LDOS.
Fig.~\ref{fig:correlated_panel}(d) shows the total exciton density of states with the
scattering-continuum and bound-sector contributions resolved separately,
confirming a discrete bound-state manifold below the continuum edge.

\begin{figure}[!t]
    \centering
    \includegraphics[]{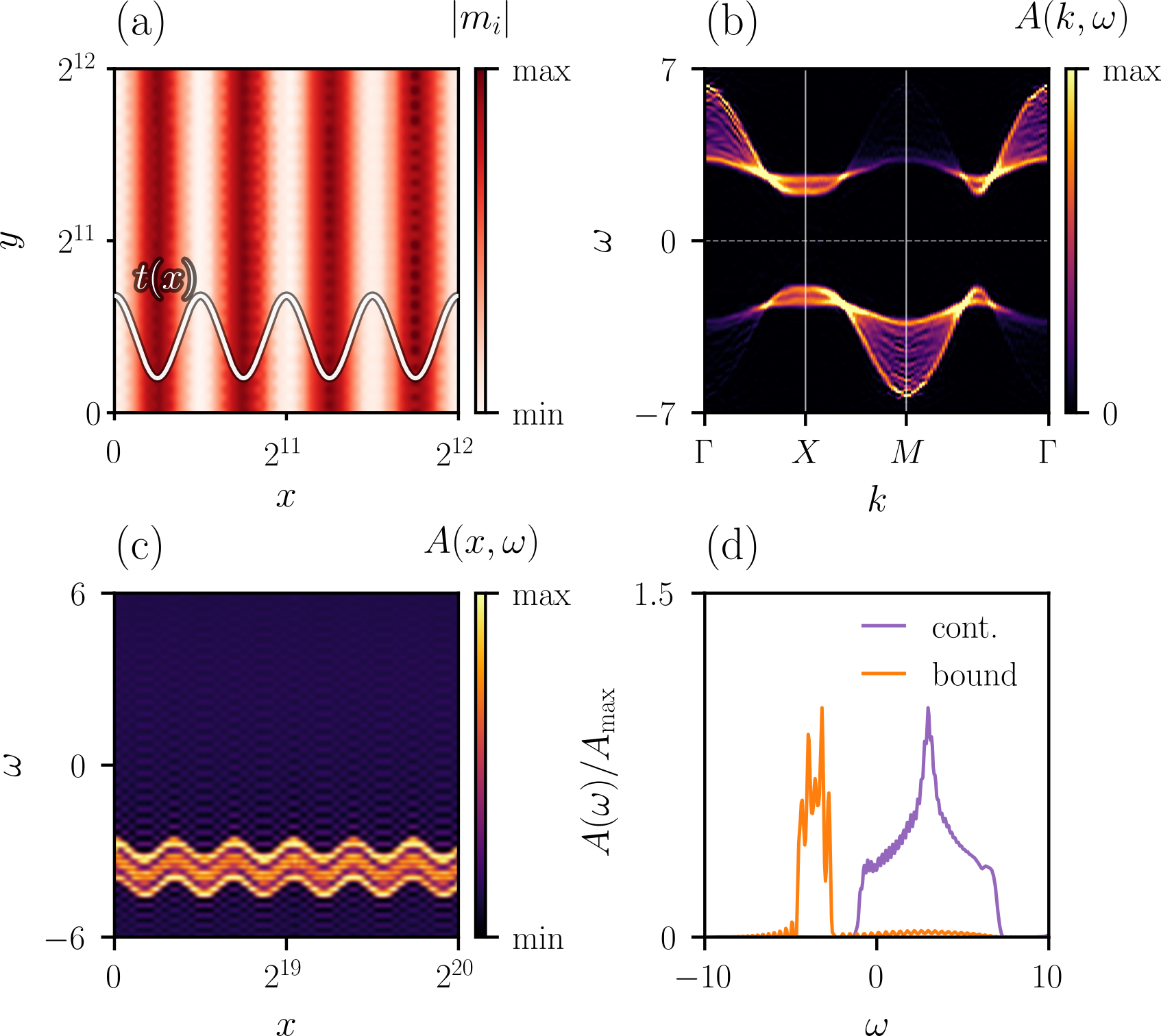}
    \caption{
Mean-field Hubbard and exciton observables for spatially modulated systems.
(a) Absolute mean-field Hubbard magnetization $|m_i|$, showing stripe order induced by the modulated hopping.
(b) Spin-summed mean-field spectral function $A(k,\omega)$ along the high-symmetry path, showing the reconstructed quasiparticle bands of the converged magnetic state and its correlated gap.
(c) Exciton local density of states $\rho(x,\omega)$, showing bound-state spectral weight localized in energy and modulated periodically in space, with moir\'e minibands as a consequence of the incommensurability of the on-site modulation $V(x)$.
(d) Total exciton density of states, separated into scattering-continuum and bound-sector contributions.
Panels (a)–(b): mean-field Hubbard calculations with $L_x=L_y=12$, $t_0=1$, hopping modulation amplitude $t_{\rm amp}=0.5$, $U=5.5$, and $N_{\rm Cheb}=100$ for the band spectrum.
Panels (c)–(d): exciton calculations with $L=20$, $t=-1$, $U=6t$, $V_0=1.5t$, scale $=10t$, $N_{\rm Cheb}=150$, HODC order $m=6$, effective broadening $\eta_{\rm eff}=1/(N_{\rm Cheb}+1-75)=1/76$, $200$ continuum and $100$ bound-sector stochastic samples.}
    \label{fig:correlated_panel}
\end{figure}

\section{Outlook}\label{sec:outlook}

Given the above discussion, it is now worth considering a series of other problems that may be suitable or unsuitable for application and further development within the tensor-network framework developed here. As we have showcased,
\pkg{TensorBinding.jl} enables
solving large-scale spatially structured models
whose Hamiltonian is compressible as a tensor network,
and it is especially tailored to rationalize physics
emerging at ultra-large length scales, such as in moir\'e, super-moir\'e, and quasicrystalline systems.
\pkg{TensorBinding.jl} thus provides a complementary set of algorithms to 
other established packages such as
Kwant \cite{groth2014}, PythTB \cite{pythtb}, Pyqula \cite{pyqula},
KITE \cite{joao2020}, or TBPLaS \cite{yuan2022}, which remain the natural choice for small-to-medium-scale
systems, transport calculations, and general sparse-matrix workflows.
On the other hand, it is worth recalling that the computational cost of the approach is controlled by the bond dimension
$\chi$ of the tensor-train representations, which grows whenever the
Hamiltonian or density matrix cannot be efficiently compressed.
Systems with dense disorder, many competing hopping ranges, or many intricate or discontinuous features can drive $\chi$ to values where the $\mathcal{O}(\chi^3)$ contraction cost becomes significant, which means that the algorithms presented here may provide less advantage over standard methods. Aditionally, the achievable spectral resolution is bounded by the number of Chebyshev
moments $N_{\rm cheb}$ and the associated KPM broadening, with finer energy
resolution requiring proportionally more iterations, and hence possible increased growth in bond dimension.
Finally, the framework is inherently single-particle, mean-field, or at most two- or few-particle in its
current form: strongly correlated many-body problems beyond Hartree-Fock, such as quantum spin liquids, fractional Chern insulators or Kondo physics lie
outside its present scope.

With this in mind, several further capabilities are currently implemented in \pkg{TensorBinding.jl} and represent active lines of development.
Topological pumps \cite{thouless1983} are accessible via cyclic adiabatic evolution of the density-matrix MPO, giving quantized charge-transport observables in periodically driven  or quasicrystalline large-scale models.
Valley Chern markers \cite{xiao2007,gilbert2025,zhang2013}, relevant to moir\'e heterostructures with broken inversion symmetry, extend the real-space topological marker framework to valley-polarized phases.
The valley projectors onto the $K$ and $K'$ points are constructed as
$\hat{P}_{K/K'} = (\mathbbm{1} \pm \hat{V})/2$,
where $\hat{V}$ is a real-space realization of the valley operator \cite{xiao2007}.
The valley Chern marker follows by encasing the Chern marker expression with the valley-projectors as
$\hat{C}_{K/K'}=P_{K/K'}\hat{C}P_{K/K'},$
providing a route for spatially resolved valley topology with
roughly the same MPO cost as the ordinary Chern marker.
Quasiparticle interference (QPI) \cite{hoffman2002}, a key observable in
scanning-tunnelling microscopy can also be obtained by placing a point-like impurity
$\hat{V}_{\rm imp} = V_0\,|\mathbf{r}_0\rangle\langle\mathbf{r}_0|$ and
computing the Fourier transform of the induced LDOS modulation,
$\delta A(\mathbf{q},\omega) = \bigl|\mathcal{F}\bigl[A_{\rm imp}(\mathbf{r},\omega)- A_{\rm clean}(\mathbf{r},\omega)\bigr]\bigr|^2,$ where both LDOS profiles $A(\mathbf{r},\omega)$ are evaluated via KPM and the
Fourier transform $\mathcal{F}$ is applied as a QFT MPO acting on the difference MPS.
Collective charge and spin excitations are captured by the RPA
\cite{bohm1953,hedin1965,aryasetiawan1998}, which resums particle-hole bubble
diagrams to all orders via
$
\hat{\chi}^{\rm RPA} = \hat{\Pi}_0(\mathbbm{1} - \hat{V}\hat{\Pi}_0)^{-1}
$ with the polarization bubble $\hat{\Pi}_0(\omega)$ assembled from a double
Chebyshev expansion in MPO arithmetic. Channels for both charge and magnetic excitations are currently implemented.
Krylov-based spectral solvers, including the Haydock recursion \cite{haydock1972}, provide an alternative to KPM, and their integration into the MPO framework is under active development. Aditional physical problems such as the compression of moir\'{e} Floquet hamiltonians, computation of transport properties in large-scale systems, and calculations related to the moiré or quasicrystalline physics of few-particle objects such as doublons and trions could represent potential research directions benefiting from our tensor network approach.
Together, these may extend \pkg{TensorBinding.jl} into a comprehensive real-space toolkit covering a wide breadth of single- and few-particle quantum problems in large-scale condensed-matter systems.

\section{Conclusion}\label{sec:conclusion}

We have presented a methodology that provides a unified strategy for using tensor network techniques to solve exceptionally large tight-binding Hamiltonians.
In particular, this enables computing properties in billion-site systems
ranging from single-particle spectral functions,
momentum-resolved band structures, real-space
topological invariants, non-Hermitian spectral responses, real-time dynamics,
self-consistent mean-field phases, and excitonic two-particle physics. 
These methodologies are implemented in \pkg{TensorBinding.jl}, an open-source Julia package for
tensor-network tight-binding calculations.
Our methodology relies on constructing large
Hamiltonians directly as tensor networks
whose bond dimensions remain modest for a broad class of
physically relevant models,
by leveraging quantics tensor cross interpolation
and tensor network algebraic algorithms.
Observables are
evaluated entirely within MPO/MPS algebra without explicit matrix
diagonalization.

\pkg{TensorBinding.jl}'s target is large-scale,
spatially structured systems where the Hamiltonian is compressible in the
quantics representation and where the relevant physics emerges at length scales
far exceeding the microscopic unit cell.
In this regime, which includes moir\'e superlattices, quasicrystals,
spatially modulated topological phases, and large exciton problems, the MPO
framework provides access to real-space observables that are otherwise
computationally prohibitive.
Together, these capabilities open a qualitatively new computational regime
for real-space condensed-matter physics, demonstrated on systems previously considered intractable: super-moir\'e structures hosting billions of atoms \cite{sun2025}, quasicrystalline Chern mosaics \cite{antao2026}, non-hermitian systems \cite{sun2026nonhermitian}, and billion-site exciton problems \cite{moustaj2026excitons}.
Looking further ahead, the MPO framework is well-positioned to address network
models of topological phase transitions, electromagnetic response of large-scale heterostructures, and real-space Floquet physics, where the relevant length scales are set by emergent structure together with the microscopic lattice.


\textbf{Acknowledgments:} We acknowledge the computational
resources provided by the Aalto Science-IT project and the
financial support from the Research Council of Finland Project
No. 370912, the
Jane and Aatos Erkko Foundation, the Finnish Quantum
Flagship, the Finnish Ministry of Education and Culture
through the Quantum Doctoral Education Pilot Program
(QDOC No. VN/3137/2024-OKM-4), the 
Finnish Quantum
Flagship (No. 358877, Aalto University), the
Finnish Centre of Excellence in Quantum Materials QMAT (No. 374166), 
InstituteQ and the ERC Consolidator
Grant ULTRATWISTROICS (Grant Agreement
No. 101170477). The authors thank the developers of ITensors.jl and QuanticsTCI.jl, on whose
infrastructure this package is built.
We also thank A. Fumega, M. Niedermeier, Q. Hoang, X. Waintal,
C. Groth, A. Akhmerov, E. V. Castro, B. Amorim, R. Oliveira, 
A. Laskowski, D. K. Morr, L. Eek, 
M. Sentef, S. Latini,
P. Torm\"a, T. Ojanen, T. Heikkil\"a, M. Johansson,
D. H. Minh Nguyen, P. Alc\'azar, G. Chen, C. Yu, P. Shen,
N. Sobrosa, P. San-Jose, T. Hansen, F. Lobo, M. Kolehmainen and J. Malag
for many useful discussions and valuable
contributions during the package's development.

\section*{Data Availability}

The source code of \pkg{TensorBinding.jl} is publicly available at
\url{https://github.com/TensorBinding/TensorBinding.jl} under an open-source
license.
All data presented in the figures were generated with the package and can be
reproduced using the large-production scripts in the folder \pkg{examples/manuscript\_files}.

\bibliography{TensorBinding}

\appendix

\section{Construction of the shift tensor}
\label{app:shift}

\begin{figure}[h]
  \centering
    \includegraphics[width=0.65\linewidth]{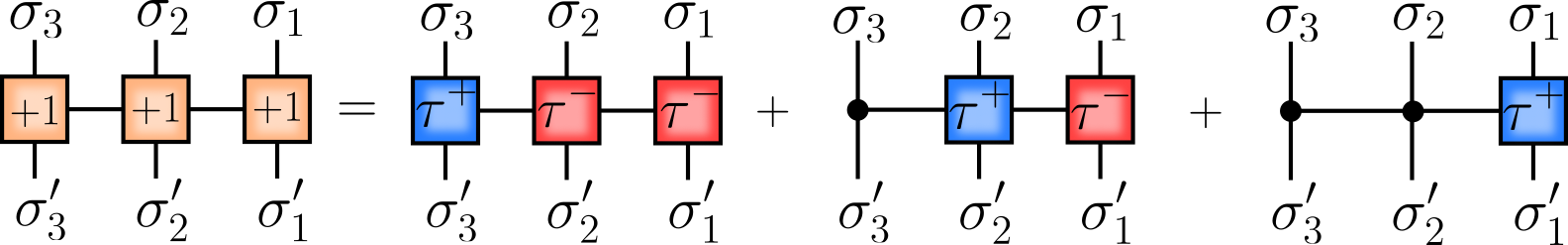}
  \caption{Shift-by-1 MPO for a three pseudospin index. Black dots represent $\delta$ or identity tensors. Colored boxes represent other tensor cores.}
  \label{fig:plus1mpo}
\end{figure}

The shift-by-$n$ MPO $\hat{S}^{(n)}$ implements $|i\rangle \mapsto |i+n \pmod{2^L}\rangle$
on an $L$-pseudospin quantics chain.
Pseudospins are ordered with site~$1$ as the least-significant bit (LSB) and site~$L$
as the most-significant bit (MSB), so the integer $i$ reads
$i = \sum_{k=0}^{L-1} \sigma_k\,2^k$ with bit $\sigma_k$ on site $L-k$. The shift-by-$1$ operator is built as a sum of tensor-product terms that
mimic binary carry propagation.
Defining $\hat\tau^+_s = |\!\uparrow\rangle\langle \downarrow\!|_s$ and
$\hat\tau^-_s = |\!\downarrow\rangle\langle \uparrow\!|_s$, the operator is
\begin{equation}
  \hat{S}^{(1)} = \sum_{k=1}^{L}\left(
    \hat\tau^+_{L+1-k}\bigotimes_{j=1}^{L-k} \hat\tau^-_j\right),
  \label{eq:opsum_shift}
\end{equation}
where the product is empty (identity) for $L-k+1$ to $L$ .
Term $k$ handles the carry at bit position $k-1$: $\hat\tau^+_{L+1-k}$ flips
the target bit from $0\to 1$, while the $\hat\tau^-_j$ operators reset
the lower bits from $1\to 0$.
Each term annihilates any state whose pseudospins do not match the required carry
pattern (since $\hat\tau^+|\!\uparrow\rangle = \hat\tau^-|\!\downarrow\rangle = 0$), so no
explicit projectors are needed. We consider an explicit example for the sake of clarity, namely for $L = 3$, Eq.~\eqref{eq:opsum_shift} expands into three terms,
\begin{align}
  \hat{T}_1 &= \mathbbm{1}_3 \otimes \mathbbm{1}_2 \otimes \hat\tau^+_1, \\
  \hat{T}_2 &= \mathbbm{1}_3 \otimes \hat\tau^+_2 \otimes \hat\tau^-_1, \\
  \hat{T}_3 &= \hat\tau^+_3 \otimes \hat\tau^-_2 \otimes \hat\tau^-_1,
\end{align}
with $\hat{S}^{(1)} = \hat{T}_1 + \hat{T}_2 + \hat{T}_3$. This MPO is illustrated in Fig.~\ref{fig:plus1mpo}.

Applying each term to $|1\rangle = |\downarrow\downarrow\uparrow\rangle$, we find
\begin{align}
  \hat{T}_1|\!\downarrow\downarrow\uparrow\rangle &= |\!\downarrow\downarrow\rangle \otimes \hat\tau^+_1|\!\uparrow\rangle_1 = 0,\nonumber\\
  \hat{T}_2|\!\downarrow\downarrow\uparrow\rangle &= |\!\downarrow\rangle_3 \otimes \hat\tau^+_2|\!\downarrow\rangle_2 \otimes \hat\tau^-_1|\!\uparrow\rangle_1
                        = |\!\downarrow\uparrow\downarrow\rangle = |2\rangle,\nonumber\\
  \hat{T}_3|\!\downarrow\downarrow\uparrow\rangle &= \hat\tau^+_3|\!\downarrow\rangle_3 \otimes \hat\tau^-_2|\!\downarrow\rangle_2 \otimes \hat\tau^-_1|\!\uparrow\rangle_1 = 0.
\end{align}
Only $\hat{T}_2$ survives: it resets bit~$0$ and carries into bit~$1$, giving $\hat{S}^{(1)}|\!\downarrow\downarrow\uparrow\rangle = |\!\downarrow\uparrow\downarrow\rangle,$ confirming that $1 + 1 = 2$.
The dense-matrix analogue of this operation is the $N\times N$ cyclic permutation matrix with $1$s
on the first sub-diagonal, and the MPO encodes the same linear map in
$\mathcal{O}(L)$ bond dimension rather than $\mathcal{O}(N^2)$ memory.

Given $\hat{S}^{(1)}$, the shift-by-$n$ operator follows by composition
$\hat{S}^{(n)} = [\hat{S}^{(1)}]^n$.
Exponentiation by squaring reduces the $n$ sequential multiplications to
$\mathcal{O}(\log_2 n)$ steps. Namely, writing $n = \sum_k \sigma_k 2^k$, one iterates replacing the accumulator $\hat{A}$ by $\hat{A}\cdot\hat{C}$ if $\sigma_k = 1$ and otherwise replacing $\hat{C}$ by $\hat{C}^2$ on the fly from the LSB to the MSB, with accumulator $\hat{A}$ initialised to the identity
and doubling register $\hat{C}$ initialised to $\hat{S}^{(1)}$.

An alternative to this construction is also provided by the magic-tensor approach \cite{waintal2026}, which
directly encodes addition by $n$ without repeated squaring.

\section{Construction of 2D lattices}
\label{app:2d}

A rectangular lattice of $N_x \times N_y = 2^{L_x} \times 2^{L_y}$ sites is
mapped to a 1D quantics chain of $L = L_x + L_y$ pseudospins via the row-major encoding 
\begin{equation}
  n(i_x, i_y) = i_x + i_y\,2^{L_x},
  \qquad
  i_x \in [0, 2^{L_x}), \quad i_y \in [0, 2^{L_y}).
  \label{eq:rowmajor_app}
\end{equation}
Starting from the right, the first $L_x$ pseudospins encode the column index $i_x$ and the last $L_y$
pseudospins encode the row index $i_y$.
A lattice displacement $(\Delta_x, \Delta_y)$ then corresponds to a shift of
\begin{equation}
  q = \Delta_x + \Delta_y\,2^{L_x}
\end{equation}
in the effective 1D chain, so the same shift-tensor machinery from
Appendix~\ref{app:shift} applies without modification.

\begin{figure}[h]
  \centering
    \includegraphics[width=0.36\linewidth]{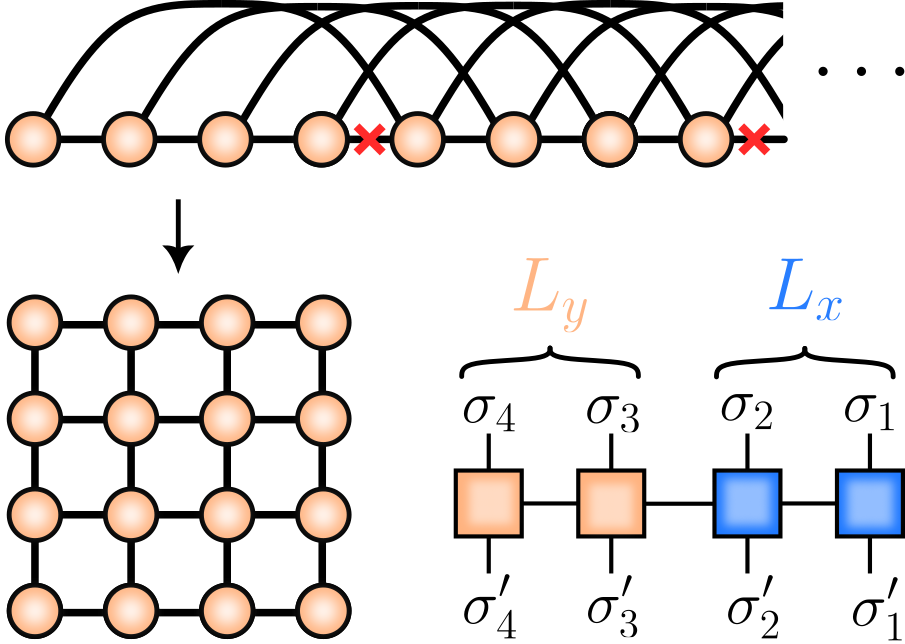}
  \caption{Row-major unfolding of a $4\times 4$ lattice ($L_x = L_y = 2$,
  $L = 4$ pseudospins) into a 1D quantics chain.
  Each site $(i_x, i_y)$ is labelled by its linear index $n = i_x + 4i_y$
  and its 4-bit quantics string $(\sigma_1\sigma_2\sigma_3\sigma_4)_2$.
  Horizontal bonds correspond to shifts $q = \pm 1$; vertical bonds to
  $q = \pm 4$.
  Red crosses mark the spurious wrap-around bonds ($n=3\to4$, $7\to8$,\dots) that must be suppressed by masking MPOs.}
  \label{fig:2d_unfolding}
\end{figure}

For $L_x = L_y = 2$ the lattice has $16$ sites and the chain has $L = 4$
pseudospins.
The linear index $n$ and the 4-bit quantics label for each site are shown in
Fig.~\ref{fig:2d_unfolding}.
The nearest-neighbour hopping directions map to the 1D shifts
\begin{alignat}{2}
  (+1,\,0) &\;\longrightarrow\; q = +1, \qquad &
  (-1,\,0) &\;\longrightarrow\; q = -1, \\
  (0,\,+1) &\;\longrightarrow\; q = +4, \qquad &
  (0,\,-1) &\;\longrightarrow\; q = -4.
\end{alignat}

It is also worth noting that the 1D chain has periodic topology, so a naive shift by $q = 1$ would connect
site $n = 2^{L_x} - 1$ (rightmost column of row $i_y$) to
$n = 2^{L_x}$ (leftmost column of row $i_y + 1$), which is not a physical
bond.
These spurious wrap-around connections are suppressed by multiplying the
hopping MPO with a masking MPO $\hat{M}$ constructed as
\begin{equation}
  \hat{M} = \mathbbm{1} - \hat{\Pi}_{\rm boundary},
\end{equation}
where $\hat{\Pi}_{\rm boundary}$ is the projector onto the offending boundary
column, built as an exact product of single-pseudospin projectors on all $L_x$
x-bit sites (sites $L_y+1$ through $L$):
\begin{equation}
  \hat{\Pi}_{i_x = N_x - 1}
  = \bigotimes_{k=L_y+1}^{L} |\!\uparrow\rangle\langle \uparrow\!|_k,
  \qquad
  \hat{\Pi}_{i_x = 0}
  = \bigotimes_{k=L_y+1}^{L} |\!\downarrow\rangle\langle \downarrow\!|_k.
\end{equation}
The first form suppresses wrap-around at the right edge of each row (needed for
forward x-hops); the second suppresses it at the left edge (needed for backward
x-hops that cross into the previous row).
Because each projector is a simple tensor product of local operators it is
exact and has bond dimension 1, adding no computational overhead.
The masked hopping term is then
\begin{equation}
  \hat{H}^{(\Delta_x,\Delta_y)}_{\rm hop}
  = \hat{M} \hat{\mathcal{T}}_{q}\,\hat{S}^{(q)} + \mathrm{h.c.}
\end{equation}
A general lattice geometry is encoded by specifying each bond type as an
integer displacement $(\Delta_x, \Delta_y)$ in the rectangular grid coordinates, which
maps to the 1D shift $q = \Delta_x + \Delta_y\cdot N_x$.
The masking required to suppress spurious wrap-arounds depends on the specific
displacement.
For a pure x-bond ($\Delta y = 0$) only a column-boundary projector is needed, as
described above.
For diagonal bonds, such as those in the triangular lattice with displacements
$(+1,+1)$ and $(-1,+1)$, a column-boundary projector alone is insufficient and
an additional row-selection mask $\hat{M}_{\rm row}$ restricts which rows
participate in each diagonal bond, preventing non-physical connections that
arise from the interleaving of the two diagonal shift directions.
For pure y-bonds ($\Delta_x = 0$) no masking is required at all, since a shift by
$q = N_x$ always maps one row to the next without crossing any column boundary.
For lattices with a basis (honeycomb, kagom\'e, Lieb, dice) the sublattice
label is carried as an auxiliary index appended to each MPO core
(see Appendix~\ref{app:aux}), and hopping amplitudes may depend on bond
direction, sublattice pair, or an arbitrary function of position compressed
via QTCI.
The same strategy extends to three dimensions by concatenating a third index
block, $n = i_x + 2^{L_x}i_y + 2^{L_x+L_y}i_z$, with masking applied
independently at each axis boundary.

\section{Construction of auxiliary degrees of freedom}
\label{app:aux}

Internal degrees of freedom such as spin, particle-hole, layer, and sublattice,
are incorporated into a position-pseudospin MPO by attaching a single extra site
carrying a $d_{\rm aux}$-dimensional auxiliary index.
This new site is connected to the existing chain with a bond of dimension 1,
so every Hamiltonian term in the extended space factorises exactly as
\begin{equation}
  \hat{H}_{\rm ext} = \hat{O}_{\rm aux} \otimes \hat{H}_{\rm pos},
\end{equation}
where $\hat{O}_{\rm aux}$ is a $d_{\rm aux}\times d_{\rm aux}$ matrix acting
on the auxiliary site and $\hat{H}_{\rm pos}$ is the original position-pseudospin
MPO.
The full Hamiltonian is then a sum of such terms with different auxiliary
matrices, and the overhead of adding the auxiliary index is exactly one extra
site at the edge of the tensor train.

\begin{figure}[h]
  \centering
    \includegraphics[width=0.5\linewidth]{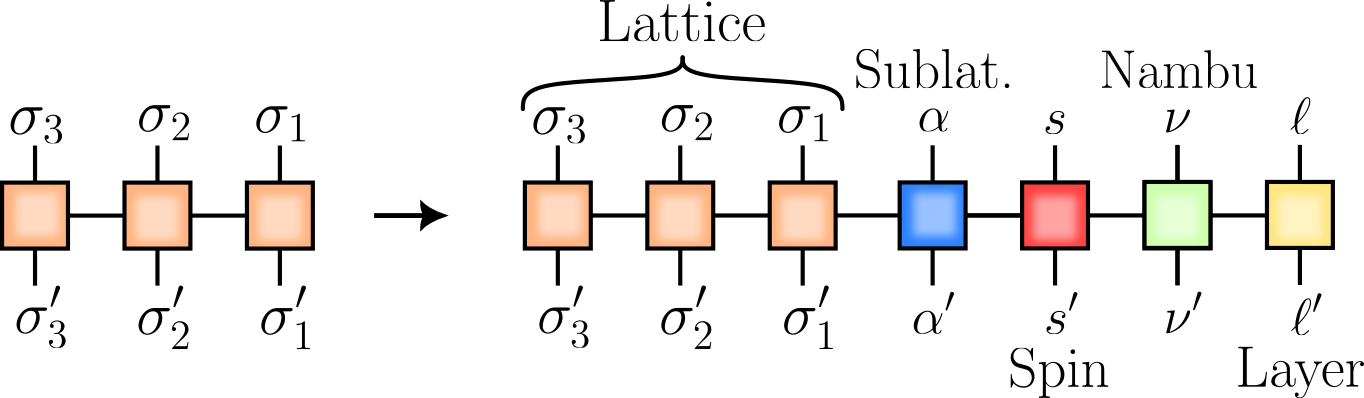}
  \caption{Example of a MPO representing a tensor network for a lattice with $N=2^3=8$ Bravais unit cells upon which is post-pended a sublattice, spin, nambu, and layer degree of freedom.}
  \label{fig:postpended}
\end{figure}

Supported auxiliary types include spin-$\tfrac{1}{2}$ ($d=2$, basis
$|\!\Uparrow\rangle, |\!\Downarrow\rangle$), Nambu electron-hole ($d=2$,
basis $|e,n\rangle\equiv c^\dagger_n|\Omega\rangle, 
|h,n\rangle \equiv c_n|\Omega\rangle$), layer indices of arbitrary dimension, and
sublattice indices (e.g.\ $d=2$ for honeycomb, $d=3$ for kagom\'e and Lieb).
Each is attached as an extra site at either edge of the tensor train. This is illustrated in terms of tensor diagrams in Fig.~\ref{fig:postpended} for the several different types of auxiliary degrees of freedom. In \pkg{TensorBinding.jl} one also has the freedom of pre- or post-pending these auxiliary tensor cores to the existing tensor trains, as different applications may benefit from LSB- or MSB-first encoding.

As a concrete illustration, starting from a position-pseudospin hopping MPO
$\hat{H}_{\rm hop}$ and on-site potential $\hat{V}$, a spin-resolved model
with a Zeeman field is assembled as
\begin{align}
  \hat{H}_{\rm kin}    &= \hat{I}_\sigma \otimes \hat{H}_{\rm hop}
    = \begin{pmatrix} \hat{H}_{\rm hop} & 0 \\ 0 & \hat{H}_{\rm hop} \end{pmatrix}, \\
  \hat{H}_{\rm Zeeman} &= \hat{S}_z \otimes \hat{V}
    = \frac{1}{2}\begin{pmatrix} \hat{V} & 0 \\ 0 & -\hat{V} \end{pmatrix},
\end{align}
where the $2\times 2$ block structure acts in spin space and each block is
itself an MPO in position space.
The full Hamiltonian $\hat{H} = \hat{H}_{\rm kin} + \hat{H}_{\rm Zeeman}$
is therefore block-diagonal in spin, with spin-up and spin-down sectors seeing
$\hat{H}_{\rm hop} \pm \tfrac{1}{2}\hat{V}$ respectively.

On the other hand, extending to a Bogoliubov-de Gennes Hamiltonian then wraps this construction
with a Nambu particle-hole layer.
The BdG Hamiltonian is assembled from three terms,
\begin{equation}
  \hat{H}_{\rm BdG}
  = \tau^z \otimes \hat{H}_{\rm hop}
  + \tau^+ \otimes \hat{\Delta}_{\rm hop}
  + \tau^- \otimes \hat{\Delta}^\dagger_{\rm hop},
\end{equation}
which in the $2\times 2$ Nambu block structure reads
\begin{equation}
  \hat{H}_{\rm BdG}
  = \begin{pmatrix}
      \hat{H}_{\rm hop} & \hat{\Delta}_{\rm hop} \\
      \hat{\Delta}^\dagger_{\rm hop} & -\hat{H}_{\rm hop}
    \end{pmatrix},
\end{equation}
where $\hat{H}_{\rm hop}$ is the normal-state kinetic MPO,
$\hat{\Delta}_{\rm hop}$ encodes the pairing amplitude, and the sign reversal
in the hole sector arises from $\tau^z = \mathrm{diag}(1,-1)$.
For a spin-resolved BdG model each block is itself a $2\times 2$ matrix in
spin space, so the full structure is a $4\times 4$ operator in the combined
Nambu-spin space with the position-pseudospin MPOs as its entries.

\section{The spectral operator using the Kernel polynomial method with tensor networks}
\label{app:KPM}

\begin{figure}[h]
  \centering
    \includegraphics[width=0.68\linewidth]{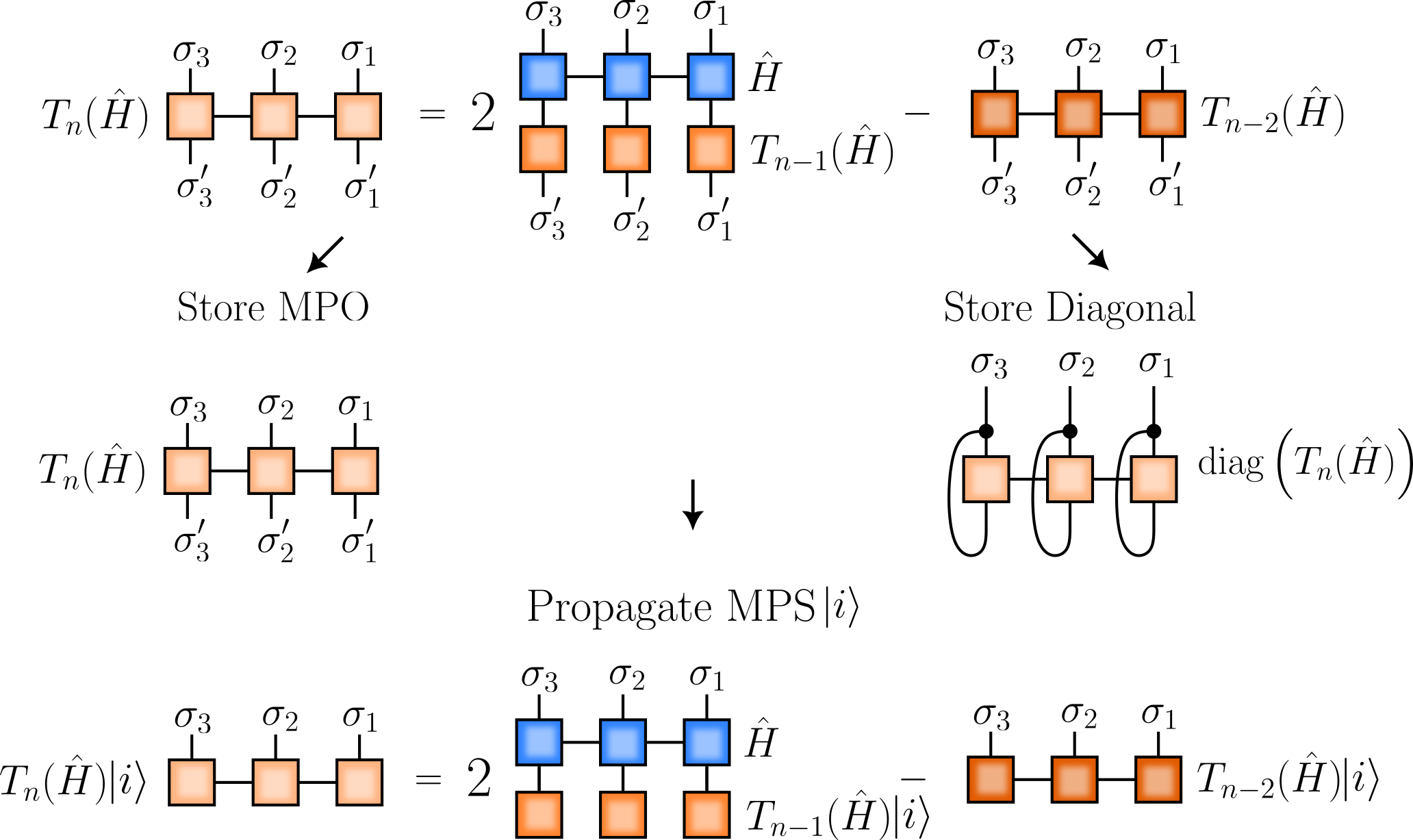}
  \caption{Tensor-network diagrams of the three KPM pathways, all built on the same Chebyshev recurrence $T_n(\hat{H}) = 2\hat{H}T_{n-1}(\hat{H})-T_{n-2}(\hat{H})$ for the rescaled Hamiltonian $\hat{H}$. At the top, the main MPO pathway, where each iteration occurs over an MPO propagated by MPO-MPO products. The left branch corresponds to storing the full MPO structure, retaining the off-diagonal spectral information.  In the right branch, the diagonal of each MPO is extracted through site-local $\delta$-tensors into an MPS encoding the site-resolved profile $A(\mathbf{r},\omega)$, with the frequency weights accumulated in a single sweep. The bottom panel showcases the MPS pathway: a reference state $|i\rangle$ is propagated by MPO-MPS products to give the MPS $T_n(\hat{H})|i\rangle$, with the moments obtained as the overlaps $\mu_n=\langle i|T_n(\hat{H})|i\rangle$.}
  \label{fig:KPMpathways}
\end{figure}

The three pathways introduced in the main text share a common kernel, namely the rescaled Hamiltonian $\hat{H}$ and the three-term Chebyshev recurrence $T_n(\hat{H}) = 2\hat{H}T_{n-1}(\hat{H})-T_{n-2}(\hat{H})$, but differ in the tensor-network object that is propagated and in how the spectral weight is finally read off. These differences are summarised diagrammatically in Fig.~\ref{fig:KPMpathways}.

For the spectral operator, the Chebyshev moments $T_n(\hat{H})$ accumulated along any of the three pathways are assembled into
\begin{equation}
  \hat{A}(\omega) \equiv \delta(\omega - \hat{H})
  \approx \sum_{n=0}^{N_c} g_n\,\hat{\mu}_n\,
  \frac{1+\delta_{n,0}}{\pi\sqrt{1-\omega^2}}\,T_n(\omega),
  \label{eq:Aw_Chebyshev}
\end{equation}
where $\hat{\mu}_n = T_n(\hat{H})$ is the $n$-th Chebyshev moment operator, $\omega\in(-1,1)$ is the energy rescaled to the Chebyshev domain, $T_n(\omega)$ is the scalar Chebyshev polynomial of the first kind evaluated at $\omega$, and $g_n$ are kernel damping coefficients that suppress Gibbs oscillations while preserving the positivity of $\hat{A}$ and the spectral sum rule. In \pkg{TensorBinding.jl}, five different kernels can be selected: the Jackson (default), Lorentz, Fejer, and Dirichlet kernels \cite{weisse2006}. Additionally, the Higher-Order Delta Chebyshev kernel \cite{Yi2025ADensities} is also available, allowing for increased spectral resolution at the cost of positivity.
The site-resolved local density of states is the diagonal matrix element of the spectral operator,
\begin{equation}
  A(\mathbf{r}_i,\omega) = \langle i\,|\,\hat{A}(\omega)\,|\,i\rangle,
  \label{eq:ldos_def}
\end{equation}
where $|i\rangle$ is the computational-basis state associated with site $i$ in the quantics encoding. Summing over all sites recovers the total density of states, $\mathrm{DOS}(\omega) = N^{-1}\sum_i A(\mathbf{r}_i,\omega) = N^{-1}\mathrm{Tr}[\hat{A}(\omega)]$. 
The three KPM pathways realise Eq.~\eqref{eq:ldos_def} in distinct ways. In the MPO pathway the full off-diagonal operator $\hat{A}(\omega)$ is materialised as an MPO at each target frequency; the LDOS at any site $i$ then follows from the scalar contraction $\langle i|\hat{A}(\omega)|i\rangle$. In the diagonal pathway both physical legs of each Chebyshev MPO iterate are contracted with a local $\delta$-tensor, projecting the operator onto its diagonal at every step; the accumulated result is an MPS encoding $A(\mathbf{r}_i,\omega)$ over all sites simultaneously, produced in a single Chebyshev sweep without ever storing the off-diagonal structure. In the MPS pathway, Eq.~\eqref{eq:ldos_def} is evaluated at a single reference site $i_0$ by choosing $|\alpha\rangle = |i_0\rangle$; the scalar moments $\mu_n = \langle i_0|T_n(\hat{H})|i_0\rangle$ are accumulated over the recurrence, and the LDOS follows from their weighted sum according to Eq.~\eqref{eq:Aw_Chebyshev}. Only three MPS are alive at any moment, making this the cheapest route when a single probe position suffices, though the diagonal pathway supersedes a loop of MPS-pathway calls since it handles all sites in one pass at comparable cost. However, the MPS pathway is the preferred choice for the two-particle Hamiltonians of Appendix~\ref{app:bse}, where the doubled chain would render the operator-valued recurrences prohibitive.

When the full spatial LDOS map $A(\mathbf{r}_i,\omega)$ is required, a choice of spatial resolution is needed. For large-scale structure, the lattice is partitioned into $n_x\times n_y$ coarse blocks of $2^a\times 2^b$ sites each, and the LDOS is reported as the block-averaged value
\begin{equation}
  \bar{A}(\mathbf{R}_p,\omega)
  = \frac{1}{2^{a+b}}\sum_{i\in\mathrm{block}\,p} A(\mathbf{r}_i,\omega),
\end{equation}
centred at the block midpoint $\mathbf{R}_p$. This sum is computed by tracing out the within-block pseudospin degrees of freedom during the MPO contraction, so the cost is independent of the block size and scales only with the number of coarse pixels. For small-scale structure, individual site values are read directly from the diagonal MPS, yielding atomic-resolution LDOS.

\section{Tensor network algorithm for momentum space spectral operators}\label{app:momentum}

Beyond real space, the spectral operator can be conjugated by the quantum Fourier transform (QFT) \cite{chen2023fourier} to give the momentum-resolved spectral function $A(\mathbf{k},\omega)$. The QFT acts on basis states $|k\rangle$ via the Fourier MPO given in Eq.~\ref{eq:Fourier_mpo} of the main text. Following the conventions established therein, the spectral function is extracted by contracting with a localised momentum-space MPS $|\mathbf{k}\rangle$,
\begin{equation}
  A(\mathbf{k},\omega)
  = \langle\mathbf{k}\,|\hat{\tilde{A}}(\omega)|\,\mathbf{k}\rangle.
  \label{eq:Akw_matrix_element}
\end{equation}
In practice, rather than constructing $\hat{\tilde{A}}(\omega)$ explicitly as an MPO sandwich at every frequency, the QFT conjugation is absorbed directly into the diagonal extraction step: the Chebyshev recurrence runs in real space, and at each step the QFT is applied to the diagonal MPS obtained from the current moment, yielding the momentum-diagonal $\langle\mathbf{k}|T_n(\hat{H})|\mathbf{k}\rangle$ as a length-$N$ vector of scalars. The weighted sum over $n$ then gives $A(\mathbf{k},\omega)$ for all $\mathbf{k}$ simultaneously, at the cost of one QFT sweep per Chebyshev moment, and the full momentum-space MPO is never stored.

This global band structure integrates spectral weight uniformly over the entire system. For spatially inhomogeneous systems, such as heterostructures, twisted bilayers, or samples hosting a defect or junction, it is useful instead to resolve both the momentum and the spatial origin of the spectral weight simultaneously. This is captured by the projected spectral function
\begin{equation}
  A_{P}(\mathbf{R},\mathbf{k},\omega)
  = \langle\mathbf{k}|\,\hat{U}_{\mathrm{QFT}}\,\hat{\mathcal{P}}_\mathbf{R}\,
    \hat{A}(\omega)\,
    \hat{\mathcal{P}}_\mathbf{R}\,\hat{U}_{\mathrm{QFT}}^\dagger\,|\mathbf{k}\rangle,
  \label{eq:Aw_proj}
\end{equation}
where $\hat{\mathcal{P}}_\mathbf{R}$ is a spatial projector centred at $\mathbf{R}$. The operator $\hat{\mathcal{P}}_\mathbf{R}$ is diagonal in the site basis, with entries given by a smooth envelope function $f_\mathbf{R}(\mathbf{r}_i)$ that is localised in a window around $\mathbf{R}$ and vanishes smoothly outside it; it is constructed as a diagonal MPO via QTCI, efficiently representing the MPO at small bond dimensions. The result is therefore a windowed and local momentum-space spectral function around $\mathbf{R}$.

\begin{figure}
    \centering
    \includegraphics[width=0.5\linewidth]{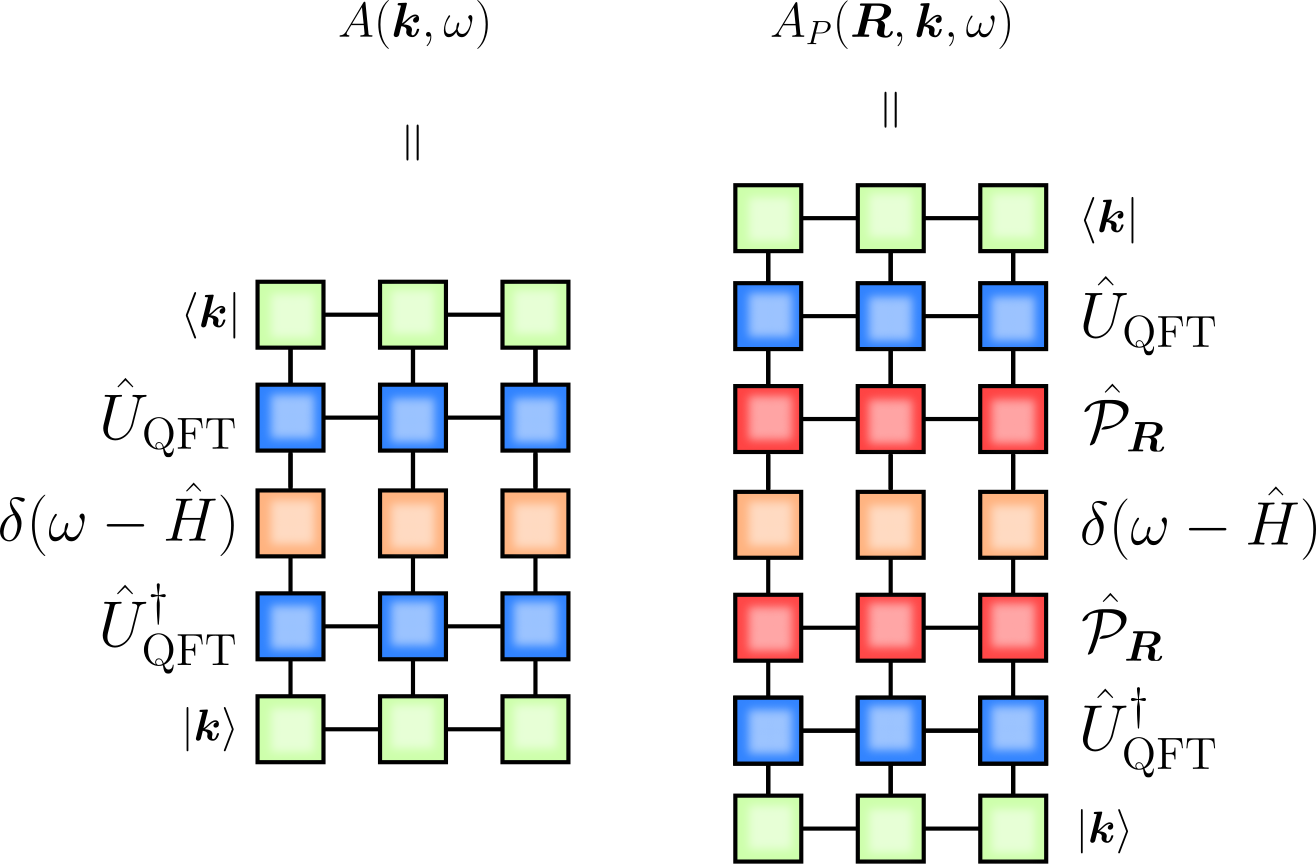}
    \caption{MPO representation of the full momentum-space spectral function $A(\mathbf{k},\omega)$ (left) and projected spectral function $A_P(\mathbf{R}, \mathbf{k},\omega)$}
    \label{fig:fourier_mpo}
\end{figure}

\section{Position-operator regularisation for real-space topological markers}
\label{app:topology}

The Chern marker and winding-number density require the position operators
$\hat{x}$ and $\hat{y}$ inside MPO expressions such as
$\hat{Q}\hat{x}\hat{P}\hat{y}\hat{Q}$.
A linear position operator has a spectrum that grows with system size, and its
MPO bond dimension likewise grows without bound: encoding $\hat{x}$ exactly on
an $L$-pseudospin chain requires bond dimension $\mathcal{O}(L)$.
In our previous work\cite{antao2026}, we find that replacing the unbounded operator with
a bounded, smooth function that agrees with it in the bulk
\begin{equation}
  \hat{x} \;\longrightarrow\; \hat{X}^\Lambda_\gamma
  \;\equiv\; \Lambda\sin\!\left(\frac{\hat{x} - x_\gamma}{\Lambda}\right),
  \label{eq:quench}
\end{equation}
where $x_\gamma$ is the physical position of the unit cell at which the marker is being evaluated and
$\Lambda$ is a length scale chosen to satisfy $\xi \ll \Lambda \ll N$, where $\xi$ is the topological correlation length set by the size of the spectral gap, and $N$ the system size.
Near the target cell, where $|x - x_\gamma| \ll \Lambda$, the sine
reduces to $x - x_\gamma$ and the topological invariant is recovered exactly.
Away from the target cell, cancellations that ensure the decay of the off-diagonal elements of the Chern operator result in no contributions to the result. Hence, because $\Lambda\sin(x/\Lambda)$ is a smooth, bounded function, its QTCI
compression into an MPO has modest bond dimension independent of $N$ and for an appropriately large choice of $\Lambda$.

\begin{figure}[h]
  \centering
    \includegraphics[width=0.5\linewidth]{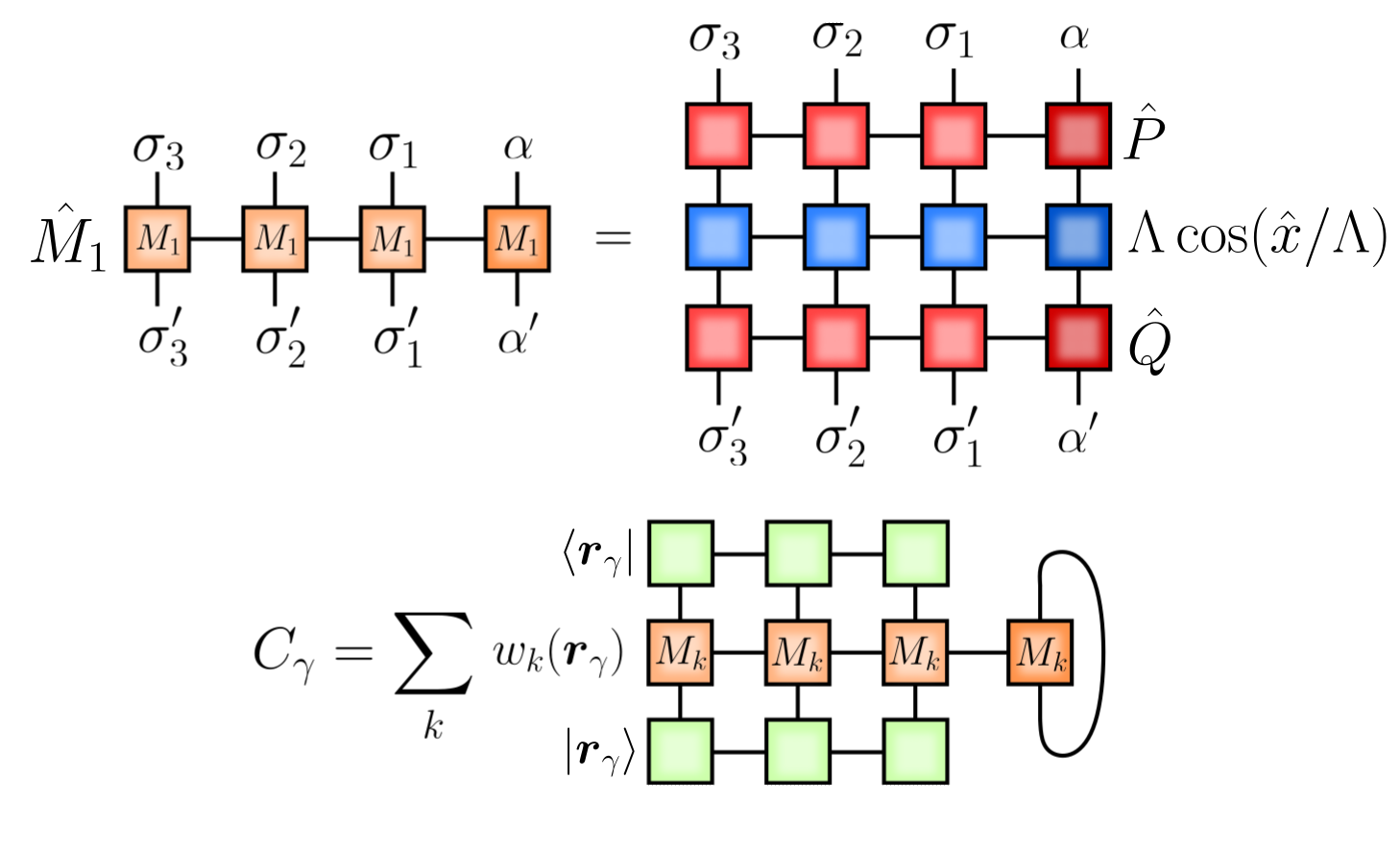}
  \caption{Tensor network algorithm utilized for the construction of a topological marker. Shaded tensors with indices labeled by $\alpha$ correspond to sublattice degrees of freedom. MPOs $\hat{M}_k$ are constructed by contracting trigonometric functions of the position operators together with occupied $\hat{P}$ and unoccupied $\hat{Q}$ state projectors. The resulting MPOs are contracted with unit cell position MPS $|\mathbf{r}_\gamma\rangle$ and the sublattice degree of freedom $\alpha$ is traced over. Finally, the topological marker at position $\gamma$ is given as a sum over the resulting scalars weighted by $w_k(\mathbf{r}_\gamma)$.}
  \label{fig:ChernMPO}
\end{figure}

It is now worth discussing the efficiency of this algorithm. A naive implementation would require
constructing a new centered operator $\hat{C}_\gamma$ for every unit cell
$\gamma$, which is prohibitive at large scales.
This is avoided by expanding the sine via the addition formula,
\begin{equation}
  \sin\!\left(\frac{\hat{x} - x_\gamma}{\Lambda}\right)
  = \sin\!\left(\frac{\hat{x}}{\Lambda}\right)\cos\!\left(\frac{x_\gamma}{\Lambda}\right)
  - \cos\!\left(\frac{\hat{x}}{\Lambda}\right)\sin\!\left(\frac{x_\gamma}{\Lambda}\right),
  \label{eq:addition}
\end{equation}
which separates the operator-valued part from the scalar weights that depend
on $\gamma$.
Substituting \eqref{eq:addition} into the Chern marker expression, one finds
that the entire marker at site $\gamma$ can be written in terms of only four
site-independent MPOs,
\begin{equation}
  \hat{C}_\gamma = \sum_{k=1}^{4} w_k(x_\gamma, y_\gamma)\,\hat{M}_k,
\end{equation}
where $\hat{M}_k$ are precomputed MPOs involving $\hat{P}$, $\hat{Q}$,
$\sin(\hat{x}/\Lambda)$, $\cos(\hat{x}/\Lambda)$,
$\sin(\hat{y}/\Lambda)$, and $\cos(\hat{y}/\Lambda)$,
and $w_k(x_\gamma,y_\gamma)$ are real scalar weights composed of
$\cos(x_\gamma/\Lambda)$, $\sin(x_\gamma/\Lambda)$,
$\cos(y_\gamma/\Lambda)$, and $\sin(y_\gamma/\Lambda)$.
For the 1D winding-number density only two such MPOs are needed.
The marker at every unit cell then follows by a summation over a trace over any existing sublattice degree of freedom along with the cheap scalar contraction $\langle\mathbf{r}_\gamma|\hat{M}_k|\mathbf{r}_\gamma\rangle$,
with no additional MPO constructions required.
This procedure, illustrated in terms of tensor network diagrams in Fig.~\ref{fig:ChernMPO} reduces the cost of a full spatial map of the topological marker to
a fixed number of
MPO precomputations regardless of how many unit cells are evaluated, namely eight for the 2D Chern marker and four for the 1D winding marker. 

\section{Algorithm for non-Hermitian KPM}
\label{app:nhkpm}

The kernel polynomial method can be generalized to non-Hermitian Hamiltonians through the hermitization described in the main text as Eq.~\ref{eq:nh_hermitization}. For a non-Hermitian tight-binding Hamiltonian $\hat{H}$, the single-particle spectral function in the complex-energy plane can be written as
\begin{equation}
f(z) = \langle \psi_L |
\delta^{(2)}(z - \hat{H}) | \psi_R \rangle ,
\end{equation}
where $\delta^{(2)}(z-z')=\delta(\Re(z)-\Re(z'))\delta(\Im(z)-\Im(z'))$ is the 2D Dirac delta function over the complex plane and
$|\psi_{L,R}\rangle$ denote the probe states used to resolve the desired
matrix element. In particular, for the local density of states, these probe
states are chosen to be localized at a lattice site. The 2D delta function can be accessed from the $z^*$ derivative
of the resolvent of the Hermitian block Hamiltonian $\tilde{H}(z)$. This gives
\begin{equation}
\hat{A}^\text{NH}(z) = \frac{1}{\pi}\,\partial_{z^*}
\langle L|\,(E - \tilde{H})^{-1}\,|R\rangle\Big|_{E=0},
\end{equation}
where the physical states are embedded in the doubled space as
$|L\rangle = |\!\uparrow\rangle_{\rm aux}\otimes|\psi_L\rangle$
and
$|R\rangle = |\!\downarrow\rangle_{\rm aux}\otimes|\psi_R\rangle$ .
Since $\tilde{H}$ is Hermitian,
its resolvent can be expanded in Chebyshev polynomials via the standard KPM
framework. 
After applying the Jackson-kernel truncation at order $N_c$, the
spectral function becomes
\begin{equation}
\hat{A}^\text{NH}(z) = \frac{2}{\pi^2}
\sum_{n=1}^{N_c} (-1)^{n+1}\,
\langle L|\,\partial_{z^*} T_{2n-1}(\tilde{H})\,|R\rangle,
\label{eq:nhkpm}
\end{equation}
where only odd-index Chebyshev polynomials $T_{2n-1}$ contribute. In addition, the $z^*$ derivative of each Chebyshev iterate obeys a modified recursion.
Differentiating the standard three-term relation $T_{n+1} = 2\tilde{H}T_n -
T_{n-1}$ with respect to $z^*$ and performing the derivative over $z^*$ on both side of the equation gives
\begin{equation}
\partial_{z^*} T_{n+1}(\tilde{H})
= \begin{pmatrix}0&0\\2&0\end{pmatrix} T_n(\tilde{H})
+ 2\tilde{H}\,\partial_{z^*}T_n(\tilde{H})
- \partial_{z^*}T_{n-1}(\tilde{H}),
\label{eq:nh_recurrence}
\end{equation}
with $\partial_{z^*}T_0=0$ and $\partial_{z^*}T_1 =
\partial_{z^*}\tilde{H}$.  The first term, proportional to $T_n(\tilde{H})$,
acts as a source that drives the derivative sequence at every step; both
the Chebyshev iterates and their $z^*$ derivatives must therefore be
propagated simultaneously.  The LDoS at site $l$ is then computed as
\begin{equation}
f(z,l) = \frac{2}{\pi^2}
\sum_{n=1}^{N_c}(-1)^{n+1}\,
\langle l\,|\mathcal{R}\,\partial_{z^*}T_{2n-1}(\tilde{H})|\,l\rangle,
\label{eq:nhkpm_ldos}
\end{equation}
where $\mathcal{R} = \mathbbm{1}\otimes\sigma^+$ projects onto the upper block,
and $|l\rangle = |\!\uparrow\rangle\otimes|l\rangle_{\rm phys}$ is the
physical site $l$ embedded in the upper half of the doubled space.
Summing over all sites yields the total non-Hermitian DOS.

The entire construction is carried out in MPO form without ever forming
the dense matrix $\tilde{H}$.  Following Appendix~\ref{app:aux}, the block
degree of freedom is introduced by attaching a single dim-$2$ auxiliary site
at either edge of the position tensor train, connected with a bond of
dimension~$1$.  Each off-diagonal block of $\tilde{H}$ factorises exactly
as $\hat{O}_{\rm NH}\otimes\hat{H}_{\rm pos}$, so the hermitized MPO reads
\begin{equation}
\hat{\mathcal{H}} =
\sigma^+\otimes(z\mathbbm{1}- \hat{H})
+
\sigma^-\otimes(z^*\mathbbm{1} - \hat{H}^\dagger),
\label{eq:nh_mpo}
\end{equation}
where $\hat{H}$ and $\mathbbm{1}$ are the non-Hermitian Hamiltonian
and identity MPOs on the position sites.  The overhead of the block index
is exactly one extra site at the edge of the tensor train, and the bond
dimension of $\hat{\mathcal{H}}$ equals that of $\hat{H}$ up to the
truncation applied when summing the two contributions.  The KPM scheme follows the third route presented in Appendix \ref{app:KPM},
iterating $T_n(\tilde{\mathcal{H}})$ and their $z^*$ derivatives as MPS; each step applies $\hat{\mathcal{H}}$ as an MPO-MPS
product followed by bond-dimension truncation.  The expectation values in
Eq.~\eqref{eq:nhkpm_ldos} are evaluated as standard MPS-MPO-MPS inner
products.

\section{RK4 algorithm for density-matrix time evolution}
\label{app:rk4}

The density matrix $\hat{\rho}(t)$ under a possibly time-dependent Hamiltonian
$\hat{H}(t)$ evolves according to the von Neumann equation
\begin{equation}
  \frac{d\hat{\rho}}{dt} = -i\bigl[\hat{H}(t),\hat{\rho}\bigr]
  = -i\bigl(\hat{H}(t)\hat{\rho} - \hat{\rho}\hat{H}(t)\bigr).
  \label{eq:vonneumann}
\end{equation}
As discussed in the main text, the density matrix is represented as an MPO on
the position-pseudospin sites, with one ket and one bra physical index at each
site.  Each application of $\hat{H}$ to $\hat{\rho}$ is therefore performed as an
MPO-MPO product, in a way such that the result is truncated immediately to control
bond-dimension growth before the next operation.

\begin{figure}[h]
    \centering
    \includegraphics[width=0.55\linewidth]{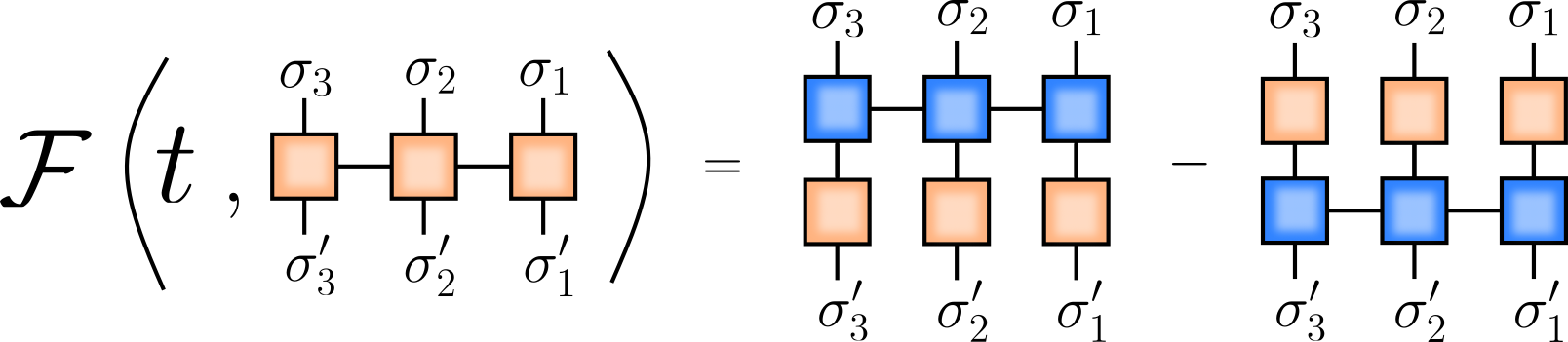}
    \caption{Action of the commutator with the MPO $\hat{H}(t)$. The target MPO is drawn in orange and $\hat{H}(t)$ in blue.}
    \label{fig:Ftmpo}
\end{figure}

Time integration uses the classical fourth-order Runge-Kutta scheme.
Writing the right-hand side of the von Neumann equation as
$\mathcal{F}(t,\hat{\rho})$, as illustrated in Fig.~\ref{fig:Ftmpo}, a single step from $\hat{\rho}(t)$ to
$\hat{\rho}(t+dt)$ evaluates four slope MPOs,
\begin{align}
  \hat{k}_1 &= \mathcal{F}(t,\; \hat{\rho}), \notag \\
  \hat{k}_2 &= \mathcal{F}\!\left(t+\tfrac{dt}{2},\;
                \hat{\rho}+\tfrac{dt}{2}\hat{k}_1\right), \notag \\
  \hat{k}_3 &= \mathcal{F}\!\left(t+\tfrac{dt}{2},\;
                \hat{\rho}+\tfrac{dt}{2}\hat{k}_2\right), \notag \\
  \hat{k}_4 &= \mathcal{F}\!\left(t+dt,\;
                \hat{\rho}+dt\cdot\hat{k}_3\right),
  \label{eq:rk4slopes}
\end{align}
and updates the density matrix as
\begin{equation}
  \hat{\rho}(t+dt) = \hat{\rho}(t)
  + \frac{dt}{6}\bigl(\hat{k}_1 + 2\hat{k}_2 + 2\hat{k}_3 + \hat{k}_4\bigr).
  \label{eq:rk4update}
\end{equation}

Each intermediate density matrix is truncated after construction.
For a time-dependent Hamiltonian the two midpoint slopes both use
$\hat{H}(t+dt/2)$, while $\hat{k}_1$ and $\hat{k}_4$ use $\hat{H}(t)$ and
$\hat{H}(t+dt)$ respectively, following the standard RK4 tableau. The local truncation error is $\mathcal{O}(dt^5)$, giving a global error
$\mathcal{O}(dt^4)$. Each slope evaluation requires two MPO-MPO products, so a full RK4 step
performs eight such contractions. 

It is also worth noticing that for a non-Hermitian $\hat{H}(t)$, the
appropriate replacement is
\begin{equation}
  \frac{d\hat{\rho}}{dt} = -i\bigl(\hat{H}(t)\hat{\rho}
  - \hat{\rho}\hat{H}(t)^\dagger\bigr),
  \label{eq:nh_vonneumann}
\end{equation}
which reduces to Eq.~\eqref{eq:vonneumann} when $\hat{H}$ is Hermitian and otherwise
allows $\mathrm{Tr}(\hat{\rho})$ to decay in the presence of gain or loss.

\section{Hartree and Fock terms as MPOs}
\label{app:hf}

Both self-energies defined in the main text are evaluated entirely within
the MPO/MPS algebra, with no need to store the full $N\times N$ matrix
elements explicitly.

\begin{figure}
    \centering
    \includegraphics[width=0.45\linewidth]{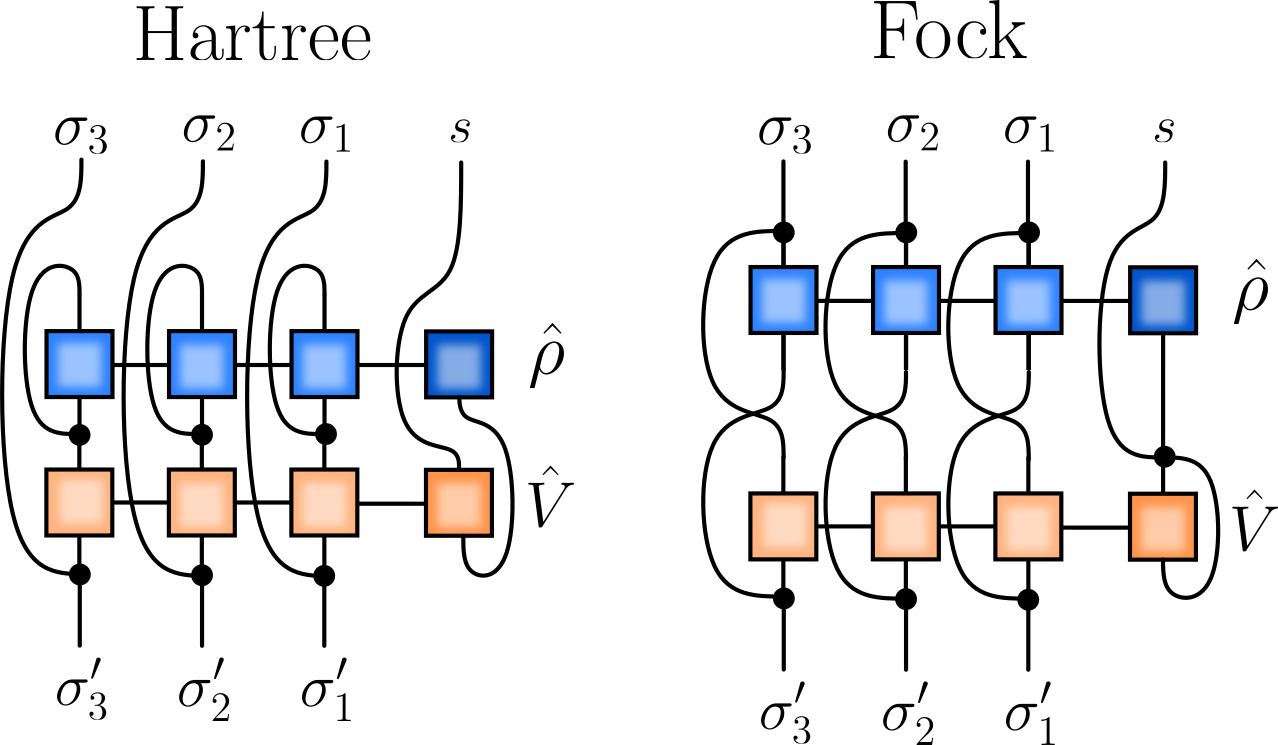}
    \caption{Hartree and Fock tensor network contractions over a quantics basis of size $N=2^3=8$ and a spin core $s$. In blue is the density matrix MPO contracted with an auxiliary spin degree of freedom. In orange is the interaction kernel MPO. black dots correspond to $\delta$ tensors that force indices to match. On the left, the Hartree term consists of the product of the diagonal elements of $\hat{\rho}$ with the MPO $\hat{V}$, resulting in an MPS which is finally promoted to a diagonal MPO. On the right, the Fock term consists of a Hadamard product between $\hat{\rho}$ and $\hat{V}$.}
    \label{fig:hfmpo}
\end{figure}

The Hartree (direct) self-energy $\Sigma^{\rm H}_{ij,s} = \delta_{ij}\sum_{k,s'}
V_{ik}^{ss'}\rho_{kk,s'}$ is diagonal in the site index and
depends on the density through the local occupations $\rho_{kk,s}$ alone.
These are extracted from the diagonal of the density-matrix MPO $\hat{\rho}$
as an MPS by reading off the diagonal element at each site.  The interaction
kernel $V_{ij}$ is constructed as an MPO using QTCI, so the sum $\sum_k V_{ik}\rho_{kk}$
is performed as an MPO-MPS product, yielding a new coefficient MPS whose
$i$-th amplitude is the Hartree potential at site $i$. This coefficient MPS
is then promoted to a diagonal MPO by tying each physical ket index to its
bra partner with a three-leg $\delta$ tensor, producing the on-site potential
$\hat{\Sigma}^{\rm H} = \mathrm{diag}(\sum_k V_{1k}\rho_k,\,
\sum_k V_{2k}\rho_k,\,\ldots)$ ready to be added to the kinetic MPO. The Hartree tensor network contraction is depicted in the left panel of Fig.~\ref{fig:hfmpo}

The Fock (exchange) self-energy $\Sigma^{\rm F}_{ij,s} =
-V_{ij}^{ss}\rho_{ij,s}$ is the element-wise product of the
interaction matrix and the full off-diagonal density matrix, and is therefore
not diagonal. It is computed as the Hadamard, or element-wise, product of the
interaction MPO $\hat{V}$ and the density-matrix MPO $\hat{\rho}$. Let us assume that these MPOs can be written as
\begin{equation}
    \hat{V}=\sum_{i,j,k\cdots,m,l}A^{(1)}_{ij}A^{(2)}_{jk}\cdots A_{ml}^{(L)}, \qquad
    \hat{\rho}=\sum_{i,j,k\cdots,m,l}B^{(1)}_{ij}B^{(2)}_{jk}\cdots B_{ml}^{(L)}
\end{equation}

At each
site $n$, the local Hadamard tensor is formed by contracting the two local MPO
cores $A^{(n)}$ (from $\hat{V}$) and $B^{(n)}$ (from $\hat{\rho}$) with a
pair of three-leg $\delta$ tensors that identify the bra index of $A$ with
the bra index of $B$ and similarly for the ket indices 
\begin{equation}
  C^{\sigma,\sigma'}_{(ij),(kl)} =
    \sum_{\sigma_n,\sigma_n'} A^{\sigma_n\sigma'_n}_{ij}\,B^{\sigma_n,\sigma_n'}_{kl}\,
    \delta_{\sigma_n,\sigma}\,\delta_{\sigma_n',\sigma'},
  \label{eq:hadamard}
\end{equation}
where the pairs $(ij),(kl)$ are the left and right virtual indices of the $A$ and $B$ tensors, grouped into two multi-indices, while $\sigma,\sigma'$ correspond to the
ket-bra physical pair of the output.  This local operation at each site produces an MPO
whose matrix elements are exactly $C^{\sigma,\sigma'} = A^{\sigma\sigma'}B^{\sigma'\sigma'}$, i.e.\ the
element-wise product, with a bond dimension that is the product of the bond
dimensions of $\hat{V}$ and $\hat{\rho}$ (followed by a truncation step).
The result, multiplied by $-1$, is the Fock MPO $\hat{\Sigma}^{\rm F}$, which
is added to the kinetic and Hartree MPOs to form the updated mean-field
Hamiltonian for the next SCF iteration. The construction of the Fock MPO in terms of tensor network contractions is given on the right panel of Fig.~\ref{fig:hfmpo}.

\section{Construction of the Bethe-Salpeter Hamiltonian}
\label{app:bse}

The central algorithmic step in constructing $\hat{H}_X$ as an MPO is the
embedding of the two independent $L$-site single-particle MPOs, $\hat{T}_c$
for the conduction electron and $\hat{T}_v$ for the valence hole, into the
interleaved $2L$-site chain.  
This ordering is introduced to preserve the locality of the electron--hole interaction and thereby reduce the computational complexity of its MPO representation. In the conventional ordering, where all electron degrees of freedom precede all hole degrees of freedom, local electron--hole interactions become long-range couplings along the tensor network, leading to an exponential growth of the MPO bond dimension with system size. In contrast, the interleaved basis places each electron pseudo-spin adjacent to its corresponding hole pseudo-spin, restoring locality and allowing the interaction operator to be represented with a constant bond dimension ($\chi=2$ for the on-site interaction considered in this work). Furthermore, this ordering naturally matches the bit-interleaved representation produced by QTCI for arbitrary interaction kernels $U(\mathbf r_e,\mathbf r_h)$, eliminating additional index permutations. Importantly, the interleaved construction is exactly equivalent to the conventional basis up to a permutation of basis states, preserving the Hamiltonian spectrum while providing a substantially more efficient tensor-network representation. 

\begin{figure}[h]
    \centering
    \includegraphics[width=0.5\linewidth]{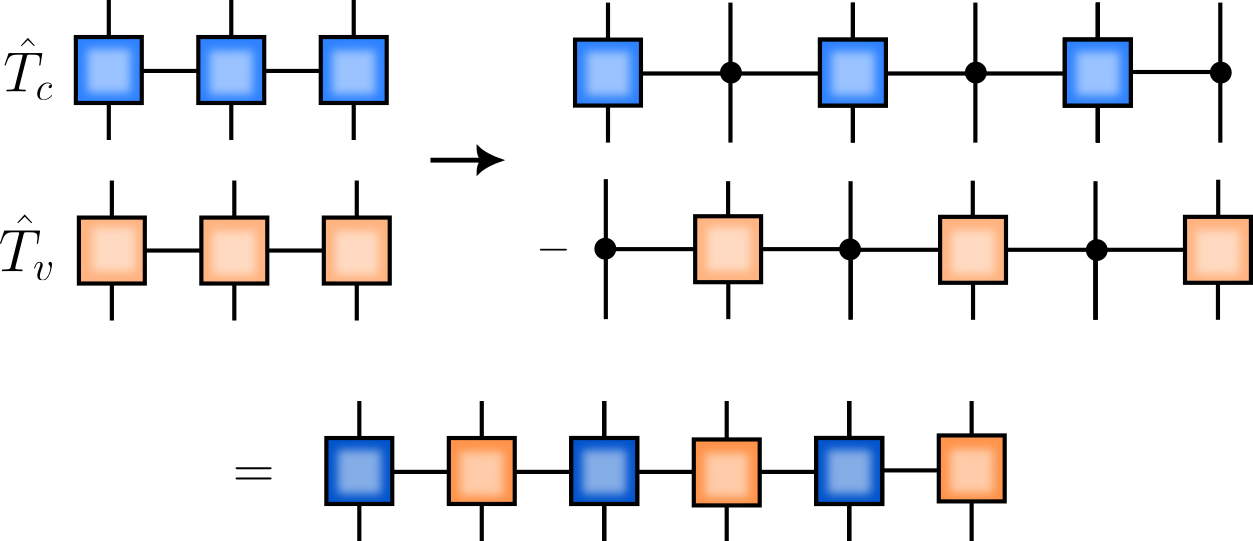}
    \caption{Algorithm for the construction of the BSE Hamiltonian as an MPO. The two independent Hamiltonians for electrons in the conduction band and holes in the valence band are first interleaved with identity or $\delta$ tensors, represented by black dots, and then subtracted from each other, resulting in the non-interacting BSE Hamiltonian.}
    \label{fig:bsempo}
\end{figure}

To construct this, each $L$-site MPO is expanded to the full
chain by inserting identity, or $\delta$ tensors at every partner site so that the
electron MPO acquires a pass-through $\delta$-tensor at each hole site, and
the hole MPO acquires one at each electron site.  The kinetic part of
$\hat{H}_X$ is then the difference of the two expanded MPOs,
$\hat{T}_c\otimes\mathbbm{1}_h - \mathbbm{1}_e\otimes\hat{T}_v$, summed on
the $2L$-site chain. This algorithm is summarized in Fig. \ref{fig:bsempo}.

The interaction $\hat{U}$, viewed as a function of the
two position labels $f(x_e, x_h) = U(x_e)\,\delta_{x_e,x_h}$, is compressed
via QTCI on the 2D $(x_e, x_h)$ grid and converted to a
diagonal MPO by contracting both physical indices on each core tensor with a
three-leg $\delta$ tensor.

\end{document}